﻿\pdfoutput=1
\documentclass[journal,onecolumn,12pt]{IEEEtran}
\usepackage[noadjust]{cite}
\usepackage{amsmath,amssymb,amsfonts}
\usepackage{algorithmic}
\usepackage{graphicx}
\usepackage{textcomp}
\usepackage{color}
\usepackage{hyperref}
\usepackage{epstopdf}
\usepackage{subfigure}
\usepackage{ntheorem}
\usepackage{booktabs,multirow}
\usepackage{enumerate}
\usepackage{tikz}
\usepackage{float}
\usepackage{epstopdf}

\newcommand{\dif}{\mathrm{d}}

\theorembodyfont{\normalfont}
\theoremheaderfont{\itshape}
\theoremseparator{:}
\newtheorem{theorem}{\hspace{2em}Theorem}

\newtheorem{definition}{\hspace{2em}Definition}
\newtheorem{lemma}{\hspace{2em}Lemma}
\newtheorem{corollary}{\hspace{2em}Corollary}
\newtheorem{remark}{\hspace{2em}Remark}
\newtheorem*{proof}{\hspace{2em}Proof}

\begin{document}
\title{Adaptive Parameter Estimation under Finite Excitation}
\author{Siyu Chen, Jing Na and Yingbo Huang
\thanks{The authors are with Faculty of Mechanical and Electrical Engineering, Kunming University of Science and Technology, Kunming, 650500, China (e-mail: chen\_siyu.acad@foxmail.com; najing25@kust.edu.cn; Yingbo Huang@126.com)}
}
\maketitle

\begin{abstract}
	Although persistent excitation is often acknowledged as a sufficient condition to exponentially converge in the field of adaptive parameter estimation, it must be noted that in practical applications this may be unguaranteed.
	Recently, more attention has turned to another relaxed condition, i.e., finite excitation.
	In this paper, for a class of nominal nonlinear systems with unknown constant parameters, a novel method that combines the Newton algorithm and the time-varying factor is proposed, which can achieve exponential convergence under finite excitation.
	First, by introducing pre-filtering, the nominal system is transformed to a linear parameterized form.
	Then the detailed mathematical derivation is outlined from an estimation error accumulated cost function.
	And it is given that the theoretical analysis of the proposed method in stability and robustness.
	Finally, comparative numerical simulations are given to illustrate the superiority of the proposed method.
\end{abstract}

\begin{IEEEkeywords}
	Adaptive parameter estimation, excitation condition, exponential convergence, time-varying gain.
\end{IEEEkeywords}

\section{Introduction}
Adaptive parameter estimation (APE) in theory has been well established throughout the past decades \cite{slotineAppliedNonlinearControl1991,krsticNonlinearAdaptiveControl1995,ioannouRobustAdaptiveControl1996}, and in application it has been used in a wide range including aerospace, chemical process and automotive systems \cite{astromSurveyAdaptiveControl1995, astromAdaptiveControl2008, eugeneRobustAdaptiveControl2013}. In some sense, the demand of APE stems from the scenario that a desired performance requires real-time tuning of the controller parameters as a result of aging, drift, or other unmanageable changes. 
Results of convergence of estimated parameters have appeared in the literature \cite{kreisselmeierAdaptiveObserversExponential1977,morganUniformAsymptoticStability1977,andersonExponentialStabilityLinear1977} where the exponential stability of APE schemes is established under a certain condition. 
With the development \cite{boydNecessarySufficientConditions1986,narendraPersistentExcitationAdaptive1987,sastryAdaptiveControlStability1990}, this condition of convergence is named as the persistent excitation (PE) condition, and gradually becomes a consensus that if it is not satisfied then APE may diverge in the presence of noise or disturbances. It is clear that the PE condition is restrictive, and may have adverse effects for sufficient.
Such as adaptive flight control, a PE reference input often results in the wastage of energy and the excessive stress on aircrafts \cite{chowdharyExponentialParameterTracking2014}.

To address this problem, various solutions have been proposed in the APE community.
Initially, the solutions were based on the idea of guaranteeing the boundedness of estimation errors without requiring the PE condition, such as leakage modifications \cite{narendraNewAdaptiveLaw1987}, projection operator \cite{lavretskyProjectionOperatorAdaptive2011} and dead zone \cite{taoAdaptiveControlDesign2003} etc.
Furthermore, some methods \cite{yucelenLowfrequencyLearningFast2013a,nguyenOptimalControlModification2012,gaudioParameterEstimationAdaptive2021} were following developed based on the aforementioned approaches to improve the weakness in transient performance.
An optimal control modification \cite{nguyenOptimalControlModification2012} was presented that can achieve robust adaptation with a large adaptive gain.
In \cite{gaudioParameterEstimationAdaptive2021}, an approach consisting of the projection operator and a time-varying learning rate has been proposed to guarantee fast parameter convergence.
However, for both approaches, the desired property of exponential convergence was not guaranteed.
Besides, an auxiliary input injection approach \cite{adetolaPerformanceImprovementAdaptive2010} has been proposed, but in many practical applications, this injection approach can lead to actuator damage, increased energy consumption, and deviations from standard operating modes.

Recently, more researchers have been turned out to study the exponential convergence under the \textit{finite/initial excitation} (FE/IE) condition, instead of the PE condition.
In \cite{aranovskiyPerformanceEnhancementParameter2017}, the authors conducted extensive research on relaxing the persistent excitation condition. The method including regressor extension and mixing, called DREM, achieved satisfactory results by a concept of assuming that the excitation horizon is finite.
In \cite{songAdaptiveControlExponential2017}, the authors proposed a new time-varying gain adaptation method, which can achieve exponential convergence of estimated parameters under full-state feedback.
Moreover, a composite model reference adaptive control framework for improved parameter convergence was proposed in \cite{choCompositeModelReference2018}, which also only requires finite excitation to converge.
Another approach \cite{glushchenkoRegressionFiltrationResetting2022} with Kreisselmeier filter and DREM is proposed to address the jump change of unknown parameters in MRAC (model reference adaptive control).

In this paper, we present a novel estimator that can achieve the exponential convergence of parameter estimation error, even though the PE condition is relaxed to the FE condition.
First, we discuss those issues encountered in currently prevalent estimators under different excitation conditions, i.e., the PE and FE conditions. Here a standard linear parameterization model is constructed by introducing the pre-filtering approach. Beginning at the cost function, a detailed step-by-step derivation is reported based on the Newton algorithm where a time-varying forgetting factor and a time-varying weight factor are introduced to ensure the boundedness of APE, culminating in the proposed Newton-based estimator (NBE).
Then, the analysis of convergence in the NBE is given, comparing with previous prevalent APE schemes discussed. And scenarios involving the presence of disturbances are also accounted for in the analysis.
Finally, comparative simulations are conducted to evaluate the theoretical studies.

The remainder of this paper is organized as follows.
Section II provides an overview of important definitions and notations, along with a review of prevalent approaches and their convergence properties.
The detailed derivation, including some key lemmas, is presented in Section III and the main theorem is further given in Sections IV and V.
Then Section VI reports the results of numerical simulations.
Section VII is the conclusion of this paper.

\noindent\textbf{Notations:}
The symbol $\|\cdot\|$ represents the 2-norm of vectors and matrices, and $I$ and 0 represent the identity matrix and a null matrix/vector with appropriate dimensions, respectively. For a matrix $A \in \mathbb{R}^{n \times n}$, $A > 0$ indicates it is positive definite (p.d.) and $A \geq 0$ indicates it is positive semi-definite (p.s.d.).
Moreover, $\mathfrak{e}_i(A)$, $i=1,2,\ldots,n$ denotes the eigenvalues of $A$. Moreover, the subscript $(\cdot)_0$ means that the associated variable is constant, and the augment $t$ involved in the notations is omitted unless necessary.

\section{Problem formulation and Preliminaries}\label{Problem formulation}

\subsection{Problem formulation}
The parameter estimation for the following system is considered:
\begin{equation}
	\label{system}
	\dot{x}(t)=\varphi(x,u) + \varPhi(x,u){\varTheta},
\end{equation}
where $x(t) \in \mathbb{R}^{q}$ is the system state with the initial value $x(t_0)=x_0$, $u(t)\in\mathbb{R}^{m}$ is the system input, $\varphi(x,u) \in \mathbb{R}^{q}$ is the known system dynamics, $\varPhi(x,u) \in \mathbb{R}^{q \times n}$ is the regressor and $\varTheta \in \mathbb{R}^{n}$ is the unknown constant parameter vector to be estimated. Without loss of generality, suppose that $\varphi(x,u)$ and $\varPhi(x,u)$ are bounded.\footnotemark\footnotetext{For the case that the regressor $\varPhi$ is unbounded, a normalization $\mathcal{Y}/\big(1+\|\varPhi_f\|^2\big) = \varPhi_f\varTheta/\big(1+\|\varPhi_f\|^2\big)$ can be used for model \eqref{lpsys}, where the main claims below are not changed \cite{gaudioParameterEstimationAdaptive2021}.}
For the ease of notation, $x(t)$ is written as $x$ and the notation $t$ in other time-dependent variables is omitted unless necessary in the subsequent developments.

The major focus of this paper is to address the parameter estimation under the following different excitation conditions.

\begin{definition}[\cite{boydNecessarySufficientConditions1986, narendraPersistentExcitationAdaptive1987,sastryAdaptiveControlStability1990, ioannouRobustAdaptiveControl1996}]
	\label{PEdif}
	Given $\alpha > 0$, if there exists any $T > 0$ such that the inequality $\int_{t}^{t+T}$ $\varPhi^{\top }(\tau)\varPhi(\tau) \dif\tau \geq \alpha I$ is established for $\forall t \geq t_0$, then the regressor $\varPhi(t)$ is said $\alpha$-persistent excitation (PE).
\end{definition}

\begin{definition}[\cite{gaudioParameterEstimationAdaptive2021}]
	\label{FEdif}
	Given $\alpha>0$, if there exists any $T > 0$ and $t_F \geq t_0$ such that the inequality $\int_{t_F}^{t_F+T}\varPhi^{\top }(\tau)\varPhi(\tau) \dif\tau \geq \alpha I$ is established, then the regressor $\varPhi(t)$ is said $\alpha$-finite excitation (FE).
\end{definition}

\begin{remark}
	Note that the notion of FE is consistent with the \textit{interval} excitation (IE) in the recent literature \cite{bobtsovGenerationNewExciting2022,panComparativeAnalysisParameter2023,korotinaNoteFixedDiscretetime2024}, i.e., the $\varPhi$ satisfying $\alpha$-FE can also be called IE over the interval $[t_F,t_F+T]$. Compared to the latter, the former omitting a determined interval emphasizes that the excitation finitely exists even though the interval is extended infinitely, such as $\varPhi(t)=e^{-t}$.
\end{remark}

\begin{remark}
	The time-dependent regressor $\varPhi(t)$ satisfies $\alpha$-PE, provided that a window $[t, t+T]$ sliding at $[t_0,+\infty)$ all imposes sufficient excitation, which indicates that there are infinite windows to retain the excitation. However, a $\alpha$-FE regressor $\varPhi(t)$ only implies that there at least exists an interval $[t_F, t_F+T]$ imposing sufficient excitation. Therefore, though the PE condition has been well recognized and used in the parameter estimation, it is stronger than the FE condition, and may be difficult to fulfill in practice. Nevertheless, the parameter estimation under the PE condition has been widely reported, while the case under the FE condition has been rarely discussed. This motivates the current study on the parameter estimation for system \eqref{system} under the FE condition.
\end{remark}

Since the derivative $\dot{x}$ in system \eqref{system} is difficult to measure directly, a well-known parameter estimation method is to design an observer for system \eqref{system} \cite{gaudioParameterEstimationAdaptive2021} and obtain the observation error $\tilde{x} = x-\hat x$ to derive an adaptive law for updating the estimated parameters $\hat \varTheta$. However, the couplings between the estimation error $\tilde{\varTheta}=\varTheta-\hat{\varTheta}$ and the observer error $\tilde x$ inevitably influence the transient convergence of such gradient algorithm based estimators \cite{ortegaModifiedParameterEstimators2020}.

To address the above problem, an alternative idea is to use the filtering operator as \cite{naRobustAdaptiveFinite2015,naRobustAdaptiveParameter2015,naReinforcedAdaptiveParameter2021} to reconstruct the parameter estimation model.
For system \eqref{system}, a low-pass filter $H(s) = 1/(ks+1)$ is applied to derive the filtered variables $x_f,\varphi_f,\varPhi_f$ as
\begin{equation}
	\left\{\begin{aligned}
		\kappa \dot{x}_{f}+x_{f}             & =x,       &  & x_f(t_0) = 0       \\
		\kappa \dot{\varphi}_f+\varphi_f     & =\varphi, &  & \varphi_f(t_0) = 0 \\
		\kappa \dot{\varPhi}_{f}+\varPhi_{f} & =\varPhi, &  & \varPhi_f(t_0) = 0 \\
	\end{aligned}\right.,
	\label{filter}
\end{equation}
where $\kappa>0$ is a manually set small constant.

Based on the Laplace transform\footnotemark\footnotetext{Here, the vanishing effects of initial values $x_0$ are ignored as \cite{bobtsovGenerationNewExciting2022}, without violating the main results.} and Eq.\eqref{filter}, one can derive a linear parameterized model $\frac{x-x_{f}}{\kappa} = \varphi_f + \varPhi_f\varTheta$, which can be rewritten as:
\begin{equation}
	\label{lpsys}
	\mathcal{Y} = \varPhi_f\varTheta,
\end{equation}
where $\mathcal{Y} \triangleq (x-x_{f})/\kappa - \varphi_f$ denotes the known system dynamics.

It is worth noting that the above filter operator maintains the excitation property of $\varPhi$ for $\varPhi_f$ in terms of Definition~\ref{PEdif} and Definition~\ref{FEdif}, which can be summarized as follows:
\begin{lemma}
	\label{lemfil}
	Considering the filter operator \eqref{filter}, if $\varPhi$ satisfies $\alpha$-FE, then $\varPhi_f$ satisfies $\alpha'$-FE with $\alpha>\alpha'>0$. 
	Furthermore, the same claim is true for $\varPhi$ satisfying $\alpha$-PE where $\varPhi_f$ satisfies $\alpha'$-PE.
\end{lemma}

\begin{proof}
	As shown in \eqref{filter}, the filter operator $H(s) = 1/(ks+1)$ is a stable, minimum phase, rational transfer function. According to \cite[Lemma~4.8.3]{ioannouRobustAdaptiveControl1996}, it is well-known that if $\varPhi$ satisfies FE or PE, the same is true for $\varPhi_f$.
\end{proof}

Although the excitation properties of $\varPhi$ and $\varPhi_f$ are similar, it should be noted that there is a mismatch $\alpha>\alpha'$ in the scale of the excitation horizon for $\varPhi$ and $\varPhi_f$, which is due to the introduced low-pass filter operator.

\subsection{Preliminary results and motivations}
For the reformulated system \eqref{lpsys}, various parameter estimation algorithms have been developed, such as gradient-based estimator (GBE \cite{ioannouRobustAdaptiveControl1996}) and least-squares estimator (LSE \cite{slotineAppliedNonlinearControl1991}). More recently, a new robust adaptive estimator (RAE \cite{naRobustAdaptiveFinite2015}) based on the estimation error has been developed, which further tailors the idea of \cite{kreisselmeierAdaptiveObserversExponential1977} by using filter operation \eqref{filter} to avoid using the derivative $\dot x$ for system \eqref{system}. All these existing results are first briefly reviewed and compared to show the motivations of this study.

\begin{table*}
	\centering
	\caption{Comparative estimation error for system \eqref{lpsys}}
	\label{varest}
	\renewcommand\arraystretch{1.5}
	\begin{tabular}{ccc}
		\hline
		\multirow{2}{*}{Excitation conditions\footnotemark} & \multicolumn{2}{c}{Estimators} \\
		\cline{2-3} & GBE, LSE, RAE & NBE \\
		\hline
		{--} & $\|\tilde{\varTheta}(t)\| \leq \|\tilde{\varTheta}(t_0)\|, t \geq t_0$ & $\|\tilde{\varTheta}(t)\| \leq \|\tilde{\varTheta}(t_0)\|, t \geq t_0$ \\
		$\alpha$-FE & $\|\tilde{\varTheta}(t)\| \leq \lambda\|\tilde{\varTheta}(t_0)\|,\lambda\in(0,1),t \geq t_F+T$ & $\lim\limits_{t\rightarrow\infty}\|\tilde{\varTheta}(t)\| =0$ \\
		$\alpha$-PE & $\lim\limits_{t\rightarrow\infty}\|\tilde{\varTheta}(t)\| =0$ & $\lim\limits_{t\rightarrow\infty}\|\tilde{\varTheta}(t)\| =0$ \\
		\hline 
	\end{tabular}
\end{table*}
\footnotetext{Here {--} denotes a general scenario where any assumptions of excitation conditions are not imposed.}

\subsubsection{Gradient-based estimator (GBE)}
The GBE is given as
\begin{equation}
	\label{GBE}
	\dot{\hat{\varTheta}} = \varGamma_0\varPhi_f^{\top}(\mathcal{Y}-\varPhi_f\hat{\varTheta}),
\end{equation}
where $\varGamma_0 > 0$ is a manually set constant learning gain. 

The basic idea in GBE is that the estimated parameters are updated by the instant predictor error, i.e., $\mathcal{Y}-\varPhi_f\hat{\varTheta} = \varPhi_f(\varTheta - \hat{\varTheta})= \varPhi_f \tilde{\varTheta}$. This estimator has a simple structure and one tuning parameter, i.e., the gain $\varGamma_0$. Due to the pure integrator form of GBE, a trade-off between the fast convergence rate with high gain and the robustness against noise with low gain should be carefully managed.
Then, the convergence properties of GBE are given as follows.

\begin{theorem}
	For the system \eqref{lpsys}, the GBE designed in \eqref{GBE} guarantees that: 
	\begin{enumerate}
		\item $\|\tilde{\varTheta}(t)\| \leq \|\tilde{\varTheta}(t_0)\|$ for $\forall t \geq t_0$;
		\item if $\varPhi(t)$ satisfies $\alpha$-FE, then $\|\tilde{\varTheta}(t)\| \leq \lambda\|\tilde{\varTheta}(t_0)\|$, $\lambda\in(0,1)$ for $t \geq t_F+T$;
		\item if $\varPhi(t)$ satisfies $\alpha$-PE, then $\lim\limits_{t\rightarrow\infty}\|\tilde{\varTheta}(t)\|=0$ exponentially.
	\end{enumerate}
\end{theorem}

\begin{proof}
	Combined with Lemma~\ref{lemfil}, it is well-known from \cite{sastryAdaptiveControlStability1990,ioannouRobustAdaptiveControl1996,taoAdaptiveControlDesign2003} that this theorem holds.
\end{proof}

\subsubsection{Least-squares estimator (LSE)}
In GBE \eqref{GBE}, the regressor product $\varPhi_f^T \varPhi_f$ with instant information is associated with the error dynamics of \eqref{GBE}, which influences the convergence of $\tilde{\varTheta}$. Thus, the learning gain $\varGamma$ should be carefully set. To address this issue such that a fast estimation is achieved, a time-varying gain is designed in the LSE algorithm as
\begin{equation}
	\label{LSE-1}
	\dot{\varGamma} = l_0\varGamma - \varGamma\varPhi_{f}^{\top}\varPhi_{f}\varGamma,
\end{equation}
where $\varGamma(t_0) = \varGamma_0 > 0$ and $l_0>0$. Then the LSE algorithm is designed as
\begin{equation}
	\label{LSE-2}
	\dot{\hat{\varTheta}} = \varGamma\varPhi_f^{\top}(\mathcal{Y}-\varPhi_f\hat{\varTheta}).
\end{equation}
In LSE \eqref{LSE-2}, the time-varying gain $\varGamma$ is obtained via \eqref{LSE-1} with the historical information of $\varPhi_{f}^{\top}\varPhi_{f}$, hence this estimator has the advantage of achieving fast convergence and averaging out the involved measurement noise \cite{slotineAppliedNonlinearControl1991}. 
But same as GBE \eqref{GBE}, the convergent properties in LSE \eqref{LSE-2} still establish on strict excitation conditions, i.e., exponential convergence under the PE condition and only bounded convergence under the FE condition.
Moreover, when only the FE condition is guaranteed, the gain $\varGamma$ in \eqref{LSE-1} will grow to infinity where $\dot{\varGamma}$ may hold $\dot{\varGamma}=l_0\varGamma>0$ because $\varGamma>0$ and $\varGamma\varPhi_{f}^{\top}\varPhi_{f}\varGamma$ is only p.s.d., resulting in potential instability \cite{slotineAppliedNonlinearControl1991}.
As summarized in \cite{ioannouRobustAdaptiveControl1996}, some simple techniques can be imposed to address this problem, such as modifying \eqref{LSE-1} as 
\begin{equation}
	\label{LSE-3}
	\dot{\varGamma} = \left\{\begin{aligned}
		&l_0\varGamma - \varGamma\varPhi_{f}^{\top}\varPhi_{f}\varGamma, & & \textrm{if} \; \|\varGamma\|\leq\bar{\gamma} \\
		&0, & & \textrm{otherwise}
	\end{aligned}\right.
\end{equation}
where $\bar{\gamma}\geq\|\varGamma_0\|$ denotes the allowable upper bound of $\varGamma$.
Then, the convergence properties for GBE designed in \eqref{LSE-2} and \eqref{LSE-3} are given as follows.\footnotemark\footnotetext{Note that although different techniques that ensure boundedness of $\varGamma$ may affect convergence rate, the convergence properties (i.e., Theorem~\ref{thelse}) are same as the standard LSE.}

\begin{theorem}\label{thelse}
	For the system \eqref{lpsys}, the LSE designed in \eqref{LSE-1} and \eqref{LSE-2} guarantees that:
	\begin{enumerate}
		\item $\|\tilde{\varTheta}(t)\| \leq \|\tilde{\varTheta}(t_0)\|$ for $\forall t \geq t_0$;
		\item if $\varPhi(t)$ satisfies $\alpha$-FE, then $\|\tilde{\varTheta}(t)\| \leq \lambda\|\tilde{\varTheta}(t_0)\|$, $\lambda\in(0,1)$ for $t \geq t_F+T$;
		\item if $\varPhi(t)$ satisfies $\alpha$-PE, then $\lim\limits_{t\rightarrow\infty}\|\tilde{\varTheta}(t)\|=0$ exponentially.
	\end{enumerate}
\end{theorem}

\begin{proof}
	Seeing \cite{liAdaptiveControlRobot1990,slotineAppliedNonlinearControl1991,ioannouRobustAdaptiveControl1996,taoAdaptiveControlDesign2003}.
\end{proof}

\subsubsection{Robust adaptive estimator (RAE)}
As shown in LSE \eqref{LSE-2}, the parameters are updated along with the instant predictor error $\varPhi_f \tilde{\varTheta}$ as the GBE \eqref{GBE}, which may be sensitive to measurement noise. To enhance the robustness by averaging the effect of instant noise involved in the instant error $\varPhi_f \tilde{\varTheta}$, inspired by the Kreisselmeier filter \cite{kreisselmeierAdaptiveObserversExponential1977}, the RAE is proposed by constructing a pair of historical information matrices:
\begin{equation}
	\label{RAE-1}
	\dot{\mathcal{M}} = -l_0\mathcal{M} + \varPhi_{f}^{\top}\varPhi_{f}, \quad \dot{\mathcal{N}} = -l_0\mathcal{N} + \varPhi^{\top}_{f}\mathcal{Y},
\end{equation}
with initial condition $\mathcal{M}(t_0) = 0$, $\mathcal{N}(t_0) = 0$ and $l_0>0$. Then, the RAE is given as
\begin{equation}
	\label{RAE-2}
	\dot{\hat{\varTheta}} = \varGamma_0 (\mathcal{N} - \mathcal{M}\hat{\varTheta}),
\end{equation}
where $\varGamma_0 > 0$ is a manually set constant learning gain. Compared with  GBE \eqref{GBE}, the historical information of regressor $\varPhi_{f}^{\top}\varPhi_{f}$ are involved in $\mathcal{M}, \mathcal{N}$, so that a fairly high adaptation gain can be used in the adaptive law \eqref{RAE-2} while retaining the robustness.

Note the low-pass filter \eqref{filter} is designed to remedy the derivative $\dot x$, and then used together with auxiliary variables $\mathcal{M}, \mathcal{N}$ as \cite{kreisselmeierAdaptiveObserversExponential1977} to construct the RAE \eqref{RAE-2} for system \eqref{system}. As reported in our previous work \cite{naRobustAdaptiveFinite2015}, the estimation error under the PE condition converges to zero even in the finite time. \footnotemark\footnotetext{To achieve finite-time convergence, one only needs to modify the adaptive law \eqref{RAE-2} as $\dot{\hat{\varTheta}} = \varGamma_0 \frac{\mathcal{M}^{\top}(\mathcal{N} - \mathcal{M}\hat{\varTheta})}{\|\mathcal{N} - \mathcal{M}\hat{\varTheta}\|}$ as \cite{naRobustAdaptiveFinite2015}.} 

Next, under the weaker FE condition, we will prove that the estimation error only converges to a compact set.

\begin{theorem}
	\label{therae}
	For the system \eqref{lpsys}, the RAE designed in \eqref{RAE-1} and \eqref{RAE-2} guarantees that:
	\begin{enumerate}
		\item $\|\tilde{\varTheta}(t)\| \leq \|\tilde{\varTheta}(t_0)\|$ for $\forall t \geq t_0$;
		\item if $\varPhi(t)$ satisfies $\alpha$-FE, then $\|\tilde{\varTheta}(t)\| \leq \lambda\|\tilde{\varTheta}(t_0)\|$, $\lambda\in(0,1)$ for $t \geq t_F+T$;
		\item if $\varPhi(t)$ satisfies $\alpha$-PE, then $\lim\limits_{t\rightarrow\infty}\|\tilde{\varTheta}(t)\|=0$ exponentially.
	\end{enumerate}
\end{theorem}

\begin{proof}
	Here, we only supplement the convergence properties under the FE condition, others reported in \cite{naRobustAdaptiveFinite2015}. 
	
	Choose a Lyapunov function $V(\tilde{\varTheta})$ as
	\begin{equation}\label{LYP}
		V(\tilde{\varTheta}) = \frac{1}{2}\tilde{\varTheta}^{\top} \tilde{\varTheta}.
	\end{equation}
	Then, $\dot{V}$ can be obtained as:
	\begin{equation}
		\dot{V} = \tilde{\varTheta}^{\top} \varGamma_0\big(\mathcal{N} - \mathcal{M}\hat{\varTheta}\big) = - \tilde{\varTheta}^{\top}\varGamma_0\mathcal{M}\tilde{\varTheta}.
	\end{equation}
	According to Lemma~\ref{lemfil}, once the regressor $\varPhi$ fulfills the FE condition, for an arbitrary vector $\omega\in\mathbb{R}^{n}$, we have
	\begin{equation}
		\label{FEtra}	
		\int_{t_F}^{t_F+T}\omega^{\top}\ell(\tau)\varPhi^{\top}_f(\tau)\varPhi_f(\tau)\omega\dif\tau\geq\alpha'.
	\end{equation}
	Then, given $t \geq t_F+T$, it follows 
	\begin{equation}
		\label{raem}
		\begin{aligned}
			\omega^{\top}\mathcal{M}(t)\omega ={}& \int_{t_0}^{t_F} e^{-l(t-\tau)} \omega^{\top}\varPhi^{\top}_f(\tau)\varPhi_f(\tau)\omega \dif\tau
			+ \int_{t_F}^{t_F+T} e^{-l(t-\tau)} \omega^{\top}\varPhi^{\top}_f(\tau)\varPhi_f(\tau)\omega \dif\tau \\
			{}&+ \int_{t_F+T}^{t_F+T} e^{-l(t-\tau)} \omega^{\top}\varPhi^{\top}_f(\tau)\varPhi_f(\tau)\omega \dif\tau \\
			\geq{}& e^{-l(t-t_F)}\int_{t_F}^{t_F+T} \omega^{\top}\varPhi^{\top}_f(\tau)\varPhi_f(\tau)\omega \dif\tau \\
			\geq{}& e^{-l(t-t_F)}\alpha',
		\end{aligned}
	\end{equation}
	i.e., given any $t'>t_F+T$, $\mathcal{M}(t)$ is p.d. for $t\in(t_F+T,t')$.
	Then, it holds that
	\begin{equation}
		V(t) = V(t_F+T) - \int_{t_F+T}^{t}\tilde{\varTheta}^{\top}(\tau)\varGamma_0\mathcal{M}(\tau)\tilde{\varTheta}(\tau)\dif\tau.
	\end{equation}
	From \eqref{raem}, one has that
	\begin{equation}
		\int_{t_F+T}^{t}\tilde{\varTheta}^{\top}(\tau)\varGamma_0\mathcal{M}(\tau)\tilde{\varTheta}(\tau)\dif\tau > 0,
	\end{equation}
	combining to $V\geq0$, so we can obtain that there must exist $\lambda\in(0,1)$ such that $\|\tilde{\varTheta}(t)\| \leq$ $\lambda\|\tilde{\varTheta}(t_F+T)\| \leq \lambda\|\tilde{\varTheta}(t_0)\|$ holds, i.e., $\tilde{\varTheta}$ converges to a compact set. 
\end{proof}

As summarized in Table~\ref{varest}, considering all the aforementioned parameter estimation algorithms, e.g., GBE, LSE and RAE, the regressor $\varPhi$ should fulfill the PE condition to retain the convergence of estimation error to zero. Under the weaker FE condition, only the boundedness of the estimation error can be retained. The purpose of this paper is to further explore the idea of \cite{naRobustAdaptiveFinite2015} using the estimation error to adapt the FE condition for retaining the estimation error convergence, and ensuring the boundedness of time-varying learning gains.

\section{Parameter Estimation under Finite Excitation}\label{Main results}
Inspired by the Newton algorithm, this section will present a new estimator to handle the parameter estimation under finite excitation for system \eqref{lpsys}. For this purpose, we define a cost function as follows:
\begin{equation}
	\label{inteindex}
	J\big(\hat{\varTheta},t\big) \triangleq \frac{1}{2} \int_{t_0}^{t} e^{-\int_\tau^t{l(r)\dif r}}\ell(\tau)\left\|\varPhi_f(\tau)\big(\varTheta-\hat{\varTheta}(t)\big)\right\|^2 \dif\tau,
\end{equation}
where $l$ acts as the forgetting factor for weighting the accumulated past error that exponentially diminishes along with time $t$, and $\ell$ is an adjustable weight. Here, $l $ and $\ell$ are all time-varying variables to be designed.

\begin{remark}
	The cost function \eqref{inteindex} is defined as the integral square of the estimation error $\tilde{\varTheta}(t)$ on all historical data, which will be minimized via the Newton algorithm to derive the adaptive law. Specifically, different to the existing results reported in \cite{liAdaptiveControlRobot1990} (i.e., the bounded-gain forgetting (BGF) algorithm), here both $l$ and $\ell$ are designed as time-varying variables, which aim to remedy the unboundedness of the gains encountered in the LSE \eqref{LSE-1} under the FE condition, and ensure the exponential convergence of estimation error over the RAE algorithm \eqref{RAE-2}.
\end{remark}

By recalling the Newton algorithm \cite{luenbergerLinearNonlinearProgramming1984,boydConvexOptimization1998,nocedalNumericalOptimization1999}, one can obtain the following estimator:
\begin{equation}
	\label{Ndis}
	\dot{\hat{\varTheta}}=-(\nabla^2 J)^{-1} \nabla J,
\end{equation}
where $\nabla J$ and $\nabla^2 J$ denote the gradient and Hessian of $J$ with respect to $\hat{\varTheta}$, which can be obtained from \eqref{inteindex} as
\begin{align}
	 & \nabla J = -\int_{t_0}^{t} e^{-\int_\tau^t{l(r)\dif r}} \ell(\tau)\varPhi^{\top}_{f}(\tau) \varPhi_{f}(\tau)\big(\varTheta-\hat{\varTheta}(t)\big) \dif\tau, \\
	 & \nabla^2 J = \int_{t_0}^{t} e^{-\int_\tau^t{l(r)\dif r}} \ell(\tau)\varPhi^{\top}_{f}(\tau) \varPhi_{f}(\tau) \dif\tau.
\end{align}

To avoid online calculating the above formulation, which is not numerically feasible since $\varTheta$ is unknown, we introduce the historical information matrices $\mathcal{M} \in \mathbb{R}^{n \times n}$ and $\mathcal{N} \in \mathbb{R}^{n}$ as
\begin{equation}
	\label{MN}
	\begin{aligned}
		\dot{\mathcal{M}} ={} & -l\mathcal{M} + \ell\varPhi_{f}^{\top}\varPhi_{f}, &  & \mathcal{M}(t_0) = 0, \\
		\dot{\mathcal{N}} ={} & -l\mathcal{N} + \ell\varPhi^{\top}_{f}\mathcal{Y}, &  & \mathcal{N}(t_0) = 0,
	\end{aligned}
\end{equation}
the solutions of which are given as follows:
\begin{equation}\label{MNslo}
	\begin{aligned}
		\mathcal{M}(t) = & \int_{t_0}^{t} e^{-\int_\tau^t{l(r)\dif r}} \ell(\tau)\varPhi^{\top}_{f}(\tau) \varPhi_{f}(\tau) \dif\tau, \\
		\mathcal{N}(t) = & \int_{t_0}^{t} e^{-\int_\tau^t{l(r)\dif r}} \ell(\tau)\varPhi^{\top}_{f}(\tau)\mathcal{Y}(\tau) \dif\tau.
	\end{aligned}
\end{equation}

Then from system \eqref{lpsys} and \eqref{MNslo}, we have:
\begin{lemma}
	\label{MNeq}
	The information matrices $\mathcal{M}$ and $\mathcal{N}$ given in \eqref{MNslo} for system \eqref{lpsys} fulfills $\mathcal{N} = \mathcal{M}{\varTheta}$.
\end{lemma}

The proof of above lemma is straightforward. Moreover, it further yields the fact
$\mathcal{M}\tilde{\varTheta} = \mathcal{N} - \mathcal{M}\hat{\varTheta}$, $\nabla^2 J = \mathcal{M}$ and $\nabla J = \mathcal{M}\hat{\varTheta} - \mathcal{N}$. Hence, with the matrices $\mathcal{M}, \mathcal{N}$ that can be online updated with \eqref{MN}, the gradient $\nabla J$ can be calculated for estimator \eqref{Ndis}. However, the inverse of Hessian, i.e., $-(\nabla^2 J)^{-1}$, is required in \eqref{Ndis}, which may not exist under the FE condition and not be numerically tractable even $\nabla^2 J = \mathcal{M}(t)$ in \eqref{Ndis} holds.

To overcome this difficulty, we will further introduce an alternative variable $\varGamma^{-1} \backsimeq \nabla^2 J$ to adapt the FE condition.
Based on the iterative form of $\mathcal{M}$ in \eqref{MN} and the equality $\dot{\varGamma} = -\varGamma\bigl(\dif\dot{\varGamma}^{-1}/\dif{t}\bigr)\varGamma$, a feasible online learning algorithm of $\varGamma$ is given by
\begin{equation}
	\label{K}
	\dot{\varGamma} = l\varGamma - \ell\varGamma\varPhi_{f}^{\top}\varPhi_{f}\varGamma, \quad \varGamma(t_0)=k_0 I > 0,
\end{equation}
where $k_0>0$ is a positive constant denoting the initial value.

The analytical solution of $\varGamma^{-1}$ is obtained as:
\begin{equation}
	\label{Kapp}
	\varGamma^{-1}(t)
	= e^{-\int_{t_0}^t{l(r)\dif r}} k^{-1}_0 I + \int_{t_0}^{t} e^{-\int_\tau^t{l(r)\dif r}} \ell(\tau)\varPhi^{\top}_{f}(\tau) \varPhi_{f}(\tau) \dif\tau
	= e^{-\int_{t_0}^t{l(r)\dif r}} k^{-1}_0 I+\nabla^2 J(t).
\end{equation}

Clearly, the first term in \eqref{Kapp} denoting the initial value $\varGamma(t_0)$ is implicitly incorporated to maintain that the time-varying gain $\varGamma$ is p.d.. Nevertheless, for sufficiently large $K_0$, it follows that $\varGamma^{-1} \rightarrow \nabla^2 J =\mathcal{M}$.

Based on the above developments, the Newton-based estimator (NBE) is designed as follows:
\begin{equation}
	\label{Newton-based}
	\dot{\hat{\varTheta}} = \varGamma\big(\mathcal{N} - \mathcal{M}\hat{\varTheta}\big),
\end{equation}
which can be implemented by using the online updated variables $\mathcal{M}, \mathcal{N}$ and $\varGamma$.

Note the essential differences between the NBE \eqref{Newton-based} and the RAE \eqref{RAE-2} with \eqref{RAE-1} is that time-varying weighting coefficients $l, \ell$ are involved in the derivation of $\mathcal{M}, \mathcal{N}$ in \eqref{MN} and a time-varying gain $\varGamma$ also with such varying coefficients in \eqref{K} is introduced in \eqref{Newton-based}. Thus, the final work is to design proper weighting coefficients $l, \ell$ to retain the boundedness of gain $\varGamma$ under the FE condition.

To ensure the boundedness, the time-varying forgetting factor $l $ is designed as
\begin{equation}
	\label{l}
	l = l_0 \big(1 - \frac{\|\varGamma\|}{\bar{\gamma}}\big),
\end{equation}
where $l_0>0$ denotes the maximum forgetting factor, and $\bar{\gamma}>k_0$ defines the allowable upper bound of $\varGamma$, i.e., $\|\varGamma\| \leq \bar{\gamma}$.

Similarly, the time-varying weight $\ell$ is designed as
\begin{equation}
	\label{ell}
	\ell = \ell_0 \big(1 - \frac{\|\mathcal{M}\|}{\bar{m}}\big),
\end{equation}
where  $\ell_0>0$ denotes the maximum adjustable weight, and $\bar{m}>0$ defines the allowable upper bound of $\mathcal{M}$, i.e., $\|\mathcal{M}\| \leq \ell_0$.

In practice, the maximum forgetting factor and adjustable weight can be set as $l_0=\ell_0=1$, and the allowable upper bounds of $\varGamma$ and $\mathcal{M}$ (i.e., $\bar{\gamma}$ and $\bar{m}$) can be tuned based on prior information of the studied system.\footnotemark\footnotetext{Note if one sets $l=l_0, \ell_0=1$ and $\varGamma=\varGamma_0$, then the NBE \eqref{Newton-based} is reduced to the RAE \eqref{RAE-2}.}

Before we give the convergence analysis of NBE \eqref{Newton-based},  the boundedness of the aforementioned intermediate variables is addressed.

\begin{lemma}
	\label{fgpro}
	Let $\bar{\gamma} > k_0$ and $\bar{m} > 0$. For $l(t)$ and $\ell(t)$ designed in \eqref{l} and \eqref{ell}, it is guaranteed that $0 \leq l(t) \leq l_0$ and $0 \leq \ell(t) \leq \ell_0$ for $\forall t \geq t_0$.
\end{lemma}

\begin{proof}
	First, it should be clarified that the range of $l$ and $\ell$ is independent of any excitation conditions.
	From the definitions given in \eqref{l} and \eqref{ell}, it is trivial to verify $\ell \leq \ell_0$ and $l \leq l_0$. Next, the properties $l \geq 0$ and $\ell \geq 0$ are proved by contradiction.
	
	According to \eqref{MN} and \eqref{ell}, one has that $\ell(t_0)=\ell_0>0$. Since all variables are continuous, we can suppose that there exists $t_s>t_0$ such that $\ell(t_s)=0$ and $\ell(t)\geq0$ for $\forall t\in[t_0,t_s]$. Moreover, given a small increment $\sigma\rightarrow0$, it is true that $\ell(t_s+\sigma)<0$, i.e., the following deduction can be established
	\begin{equation}
		\label{lim1}
		\lim\limits_{\sigma\rightarrow0}\frac{\ell(t_s+\sigma)}{\sigma} < 0.
	\end{equation}
	Substituting \eqref{MN} and \eqref{ell} into \eqref{lim1}, it is derived that
	\begin{equation}\label{ellg}
		\begin{split}
			\lim\limits_{\sigma \rightarrow 0}\frac{\ell(t_a+\sigma)}{\sigma} ={}& \lim\limits_{\sigma \rightarrow 0}\frac{\ell_0\Bigl(1-\frac{\|\mathcal{M}(t_s+\sigma)\|}{\bar{m}}\Bigr)}{\sigma} \\
			={}& \lim\limits_{\sigma \rightarrow 0}\frac{\ell_0\Bigl(1-\frac{\|\mathcal{M}(t_s)+\sigma\dot{\mathcal{M}}(t_s)\|}{\bar{m}}\Bigr)}{\sigma} \\
			={}& \lim\limits_{\sigma \rightarrow 0}\frac{\ell_0\Bigl(1-\bigl(1-\sigma l(t_s)\bigr)\frac{\|\mathcal{M}(t_s)\|}{\bar{m}}\Bigr)}{\sigma} \\
			={}& \ell_0l(t_s).
		\end{split}
	\end{equation}
	Hence it is concluded that $\ell(t_s+\sigma)<0$ exists if and only if $l(t_s)<0$.

	Next, based on \eqref{MNslo} and $\ell(t)\geq0$, it is clear that $\mathcal{M}(t)$ is p.s.d. for $t\in[t_0,t_s]$.
	Substituting \eqref{l} into \eqref{Kapp}, it follows that
	\begin{equation}
		\label{Kff}
		\frac{\dif }{\dif t} \big(\varGamma^{-1}\big) = -l_0\varGamma^{-1} + \frac{l_0}{\bar{\gamma}}\|\varGamma\|\varGamma^{-1} + \ell\varPhi_f^{\top}\varPhi_f.
	\end{equation}
	Since it is shown $\ell(t) \geq 0$ and $\mathcal{M}(t) \geq 0$ for $t\in[t_0,t_s]$, and the inequality $\|\varGamma\|\varGamma^{-1} \geq I$ holds, we have
	\begin{multline}\label{gamaa}
		\varGamma^{-1}(t_s) = \varGamma^{-1}(t_0)e^{-l_0 t_s} +
		\int_{t_0}^{t_s}e^{-l_0 (t_s-\tau)}\Bigg(\frac{l_0}{\bar{\gamma}}\|\varGamma(\tau)\|\varGamma^{-1}(\tau) + \ell(\tau)\varPhi_f^{\top}(\tau)\varPhi_f(\tau)\Bigg) \dif \tau \\
		\geq k'I + \int_{t_0}^{t_s}e^{-l_0 (t_s-\tau)}\ell(\tau)\varPhi_f^{\top}(\tau)\varPhi_f(\tau)\dif\tau \geq k'I,
	\end{multline}
	where $k' \triangleq e^{-l_0 t_s}(k_0^{-1}-\bar{\gamma}^{-1}) + \bar{\gamma}^{-1}\geq \bar{\gamma}^{-1}$.
	Hence, it can be concluded that $\varGamma(t_s) \leq \bar{\gamma}I$, and then $l(t_s) \geq 0$, which is conflict with the above result \eqref{lim1}, i.e., $\ell(t) < 0$ cannot exist.
	Furthermore, based on \eqref{gamaa}, it is true that $\varGamma(t_s) \leq \bar{\gamma}I$ (i.e., $l(t_s) \geq 0$) for $\forall t\geq t_0$.
\end{proof}

Following the above discussions, we will further prove the boundedness of learning gain $\varGamma$ and the information matrices $\mathcal{M},\; \mathcal{N}$.

\begin{corollary}
	\label{corGamma}
	Supposing that $\varPhi$ is bounded, the time-varying gain $\varGamma$ in \eqref{K} is bounded for any $t \geq t_0$.
\end{corollary}

\begin{proof}
	This can be trivially verified from \eqref{K} with \eqref{l} and \eqref{ell}, and given in \eqref{gamaa} in the proof of Lemma~\ref{fgpro}.
\end{proof}

\begin{corollary}
	\label{corMN}
	Supposing that $\varPhi$ is bounded, the information matrices $\mathcal{M}$ and $\mathcal{N}$ in \eqref{MN} are bounded for any $t \geq t_0$.
\end{corollary}
\begin{proof}
	The boundedness of $\mathcal{M}$ can be verified from \eqref{MNslo} for bounded $l$ and $\varPhi$. Moreover, the boundedness of $\mathcal{N}$ can be verified from Lemma~\ref{MNeq} and the boundedness of $\mathcal{M}(t)$.
\end{proof}

\begin{remark}\label{remark infinite gain}
	The main contribution is to design time-varying variables $l$ and $\ell$ as given in \eqref{l} and \eqref{ell}, which can guarantee the boundedness of $\mathcal{M}$ and $\varGamma$ even under the FE condition. In fact, the LSE with a constant $l_0$ for $\varGamma$ in \eqref{LSE-1} and the RAE with constant $l_0$ and $\varGamma_0$ may lead to unbounded increases in $\mathcal{M}$ and $\varGamma$.
	Illustrative examples are provided to showcase this issue. First, suppose $l(t)=1$ is a constant and the regressor of the studied system defined is $\varPhi(t) = 1, t \in [t_0,t_c]$ and $\varPhi(t) = 0, t \in (t_c,\infty)$. Then, we know $\varPhi_f(t) = 1-e^{-\frac{t}{\kappa}}$ from \eqref{filter} which does not fulfill the PE condition, such that $\varGamma(t) = \varGamma(t_c)e^{t}$ increases to infinity according to \eqref{K}. Second, for any $\varPhi(t) \neq 0$, a constant $\ell=1$ and a time-varying $l$ as \eqref{l} are used. Once $\varGamma$ increases to reach its upper bound (i.e., $\|\varGamma\|=\bar{\gamma}$) under the FE condition, then $l=0$ is true from \eqref{l}, then $\dot{\mathcal{M}} = \varPhi_{f}^{\top}\varPhi_{f}$ is true from \eqref{MN}, which indicates that $\mathcal{M}$ may increases to infinite. However, with time-varying coefficients given in \eqref{l} and \eqref{ell}, both $\mathcal{M}$ and $\varGamma$ are bounded as given in Corollary~\ref{corGamma} and Corollary~\ref{corMN}, which also contribute to retain the convergence of the NBE in \eqref{Newton-based}. Although other existing approaches, such as the projection operator \cite{gaudioParameterEstimationAdaptive2021}, the leakage factor \cite{guayTimevaryingExtremumseekingControl2015} and time-varying forgetting factor \cite{slotineAppliedNonlinearControl1991}, can also be used to retain the boundedness of gain $\varGamma$, they cannot adapt to the FE condition to retain the convergence for the estimator.
\end{remark}

\section{Convergence Analysis}\label{Convergence Analysis}

The main results on the convergence for the NBE \eqref{Newton-based} are given as follows.
\begin{theorem}
	\label{main}
	For reformulated model \eqref{lpsys} of system \eqref{system}, the NBE designed in \eqref{Newton-based} can guarantee:
	\begin{enumerate}
		\item $\|\tilde{\varTheta}(t)\| \leq \|\tilde{\varTheta}(t_0)\|$ for $\forall t \geq t_0$;
		\item if $\varPhi(t)$ satisfies $\alpha$-FE (or $\alpha$-PE), then $\lim\limits_{t\rightarrow\infty}\|\tilde{\varTheta}(t)\|=0$ exponentially.
	\end{enumerate}
\end{theorem}

\begin{proof}
	1) We choose a Lyapunov function $V(\tilde{\varTheta}) = \frac{1}{2}\tilde{\varTheta}^{\top} \tilde{\varTheta}$ as \eqref{LYP}.
	Then from Lemma~\ref{MNeq} and \eqref{Newton-based}, the derivative of $V(\tilde{\varTheta})$ can be calculated as:
	\begin{equation}
		\label{lyav1}
		\dot{V} = - \tilde{\varTheta}^{\top}\varGamma\mathcal{M} \tilde{\varTheta} \leq -\mathfrak{e}_{\min}\big(\varGamma\mathcal{M}\big)\|\tilde{\varTheta}\|^2.
	\end{equation}

	Now, we will address the positive definiteness of $\mathfrak{e}_{\min}\big(\varGamma\mathcal{M}\big)$. 
	From the definition of $\mathcal{M}$ given in \eqref{MN}, it is symmetric and non-negative.
	Hence, by using the singular value decomposition (SVD), it can be further reformulated as $\mathcal{M}=U\mathcal{G}U^{\top}$ where $\mathcal{G}= \textnormal{diag}(\mathfrak{e}_1(\mathcal{M}), \cdots, \mathfrak{e}_n(\mathcal{M}))\in\mathbb{R}^{n\times n}$ is a diagonal matrix and $U\in\mathbb{R}^{n\times n}$ is the unitary matrix. 
	Then, we obtain from \eqref{Kapp} that $\varGamma = U\big(e^{-\int_{t_0}^t{l(r)\dif r}} k^{-1}_0 I + \mathcal{G}\big)^{-1} U^{\top}$. 
	The matrix product $\varGamma\mathcal{M}$ is further derived as $\varGamma\mathcal{M}= U\mathcal{D}U^{\top}$, where $\mathcal{D}$ is a diagonal matrix with the eigenvalues as follows:
	\begin{equation}
		\mathcal{D} = \big(e^{-\int_{t_0}^t{l(r)\dif r}} k^{-1}_0 I + \mathcal{G}\big)^{-1}\mathcal{G}
		= \textnormal{diag}\Bigg(\frac{\mathfrak{e}_1(\mathcal{M})}{\mathfrak{e}_1(\mathcal{M})+e^{-\int_{t_0}^t{l(r)\dif r}} k^{-1}_0},\cdots,
		\frac{\mathfrak{e}_n(\mathcal{M})}{\mathfrak{e}_n(\mathcal{M})+e^{-\int_{t_0}^t{l(r)\dif r}} k^{-1}_0}\Bigg).
	\end{equation}

	Therefore, $\mathcal{\varGamma\mathcal{M}}$ is also non-negative, which indicates that $\dot{V} \leq 0$ always holds. According to the Lyapunov Theorem, this proves Theorem~\ref{main}-1).

	2) Next, we will consider the case of that $\varPhi$ satisfies $\alpha$-FE. Let $\bar{m} > \|\mathcal{M}\|$ holds for $t\in[t_0,t_F+T]$, which indicates that $\ell > 0$ for $t\in[t_0,t_0+T]$. 
	For an arbitrary unit vector $\omega\in\mathbb{R}^{n}$, it is similar to \eqref{raem} that
	\begin{multline}
		\label{M_FE}
		\omega^{\top}\mathcal{M}(t)\omega = \int_{t_0}^{t_F} e^{-\int_\tau^t{l(r)\dif r}} \omega^{\top}\ell(\tau)\varPhi^{\top}_f(\tau)\varPhi_f(\tau)\omega \dif\tau \\
		+ \int_{t_F}^{t_F+T} e^{-\int_\tau^t{l(r)\dif r}} \omega^{\top}\ell(\tau)\varPhi^{\top}_f(\tau)\varPhi_f(\tau)\omega \dif\tau
		+ \int_{t_F+T}^{t} e^{-\int_\tau^t{l(r)\dif r}} \omega^{\top}\ell(\tau)\varPhi^{\top}_f(\tau)\varPhi_f(\tau)\omega \dif\tau \\
		\geq e^{-\int_{t_F}^{t_F+T}{l(r)\dif r}}\ell_F \int_{t_F}^{t_F+T} \omega^{\top}\varPhi^{\top}_f(\tau)\varPhi_f(\tau)\omega \dif\tau
		\geq e^{-\int_{t_F}^{t_F+T}{l(r)\dif r}} \ell_F\alpha'
	\end{multline}
	where $\ell_F\triangleq\inf_{t_F \leq t \leq t_F+T} \ell(t) > 0$.
	Then, it is easy to know that
	\begin{equation}
		\label{Dpos}
		\mathcal{D} \geq \frac{\mathfrak{e}_{\min}(\mathcal{M})}{\mathfrak{e}_{\min}(\mathcal{M}) + e^{-\int_{t_0}^t{l(r)\dif r}} k^{-1}_0} I
		= \frac{e^{-\int_{t_0}^t{l(r)\dif r}} \ell_F\alpha'}{e^{-\int_{t_0}^t{l(r)\dif r}} \ell_F\alpha'+e^{-\int_{t_0}^t{l(r)\dif r}} k^{-1}_0} I = \beta_F I
	\end{equation}
	where $\beta_F\triangleq\alpha'\ell_F/(\alpha'\ell_F+k^{-1}_0)>0$. 	Therefore, we can conclude that $\varGamma\mathcal{M}$ is p.d. even under the FE condition. 
	Then, it is clear that \eqref{lyav1} can be reduced to $\dot{V} \leq -\mathfrak{e}_{\min}\big(\varGamma\mathcal{M}\big)\|\tilde{\varTheta}\|^2 \leq -\varpi V$ for a positive constant $\varpi$
	Thus, it can be concluded that $\tilde{\varTheta}$ exponentially converges to zero for $t \geq t_F+T$. 

	Furthermore, if $\varPhi$ satisfies $\alpha$-PE, from the solution of $\mathcal{M}$ in \eqref{MNslo}, one also has that
	\begin{multline}
		\label{M_PE}
		\omega^{\top}\mathcal{M}(t)\omega = \\
		\int_{t_0}^{t-T} e^{-\int_{\tau}^t{l(r)\dif r}}\omega^{\top}\ell(\tau)\varPhi^{\top}_f(\tau)\varPhi_f(\tau)\omega \dif\tau
		+ \int_{t-T}^{t} e^{-\int_{\tau}^t{l(r)\dif r}}\omega^{\top}\ell(\tau)\varPhi^{\top}_f(\tau)\varPhi_f(\tau)\omega \dif\tau \\
		\geq e^{-\int_{t-T}^t{l(r)\dif r}}\ell_P\int_{t-T}^{t} \omega^{\top}\varPhi^{\top}_f(\tau)\varPhi_f(\tau)\omega \dif\tau
		\geq e^{-\int_{t-T}^t{l(r)\dif r}} \ell_P\alpha',
	\end{multline}
	with $\ell_P\triangleq\inf_{t-T \leq t} \ell$. 
	Substituting \eqref{M_PE} into \eqref{Dpos}, it derives that $\varGamma\mathcal{M} \geq \beta_P I$ where $\beta_P\triangleq\alpha'\ell_P/(\alpha'\ell_P$ $+e^{-\int_{t_0}^{t-T}{l(r)\dif r}} k^{-1}_0)$, i.e., $\varGamma\mathcal{M}$ is p.d.. 
	As same as the above result of FE, it can be concluded that $\lim\limits_{t\rightarrow\infty}\|\tilde{\varTheta}(t)\|=0$ exponentially. This proves Theorem~\ref{main}-2).
\end{proof}

\begin{remark}
	Essentially different to the GBE, the LSE and the RAE, for the proposed NBE, the weak FE condition is sufficient to guarantee the positive definiteness of $\mathcal{M}$ and thus ensure the exponential convergence of $\tilde{\varTheta}$. In fact, following the inequality \eqref{Dpos}, we have $\mathcal{D}\geq \beta_F I$ under the FE condition, such that $\varGamma\mathcal{M}>0$. Relaxation of the PE condition to the FE condition can also be interpreted by their excitation horizons. Indeed, from \eqref{M_PE} and \eqref{M_FE},  we see that in the case of PE, the excitation horizon is dependent on an integral of finite interval $e^{-\int_{t-T}^t{l(r)\dif r}}$, while under the FE condition it is determined by an infinite one where the function $l$ is square integrable.
\end{remark}

\begin{remark}
	According to the argument in \cite[Chapter 3]{nocedalNumericalOptimization1999,dennisQuasiNewtonMethodsMotivation1977}, the NBE \eqref{Newton-based} is indeed a quasi-Newton algorithm as a result of $\varGamma^{-1} \backsimeq \nabla^2 J$. Compared to the gradient-based method, the convergence rate of quasi-Newton algorithm is faster, i.e., superlinearly. This property is guaranteed if the Hessian $\nabla^2 J$, where it is $\mathcal{M}$, satisfies the Lipschitz continuous and second-order sufficient conditions. According to the discussions on Theorem~\ref{main}, both the preassumptions are satisfied under the required excitation conditions holding. Nevertheless, once $\mathcal{M}$ is p.d., the cost function $J$ attains its minimum only when $\hat{\varTheta}(t)=\varTheta$. The Lipschitz continuity assumption is equivalent to the existence of $m>0$ such that $\varGamma^{-1}\backsimeq \nabla^2 J \preceq mI$.
\end{remark}

\section{Robustness Analysis}\label{Robust Analysis}
For robustness of the proposed NBE, consider the system \eqref{system} subject to external disturbances reformulated as
\begin{equation}\label{system-d}
	\dot{x}(t)=\varphi(x,u) + \varPhi(x,u){\varTheta} + \Delta
\end{equation}
where $\Delta$ denotes bounded modeling uncertainties and measurement noise. Hence, Eq.\eqref{lpsys} can be reformulated as
\begin{equation}\label{psystem-d}
	\mathcal{Y} = \varPhi_f\varTheta + \Delta_f, \quad \kappa\dot{\Delta}_f + \Delta_f = \Delta
\end{equation}

Then, the estimation error for the NBE \eqref{Newton-based} can be represented as
\begin{equation}
	\begin{aligned}
		\label{NBE_d}
		\dot{\tilde{\varTheta}} & = - \varGamma \bigl(\mathcal{M}\tilde{\varTheta} + \varLambda\bigr)
	\end{aligned}
\end{equation}
where the residual term $\varLambda(t) = \int_{0}^{t}e^{-\int_\tau^t{l(r)\dif r}} \ell(\tau)\varPhi_f(\tau)\Delta_f^{\top}(\tau)\textnormal{d}\tau$ is bounded as long as $\Delta$ is bounded.

\begin{theorem}
	For the system given by \eqref{system-d} and \eqref{psystem-d} with bounded disturbances $\Delta$, the NBE \eqref{Newton-based} guarantees that $\tilde{\varTheta}$ converges to a compact set.
\end{theorem}

\begin{proof}
	Choose a Lyapunov function as $V(\tilde{\varTheta}) = \frac{1}{2}\tilde{\varTheta}^{\top} \tilde{\varTheta}$ as \eqref{LYP}.
	From \eqref{NBE_d}, we can derive that
	\begin{equation}
		\dot{V} = - \tilde{\varTheta}\varGamma \bigl(\mathcal{M}\tilde{\varTheta} + \varLambda\bigr) \leq -\mathfrak{e}_{\min}(\varGamma\mathcal{M})\|\tilde{\varTheta}\|^2 - \tilde{\varTheta}\varGamma\varLambda.
	\end{equation}

	Applying the Young's inequality $\pm a^{\top}b \leq a^{\top}a/2\eta+\eta b^{\top}b/2$ on the final term, then we have
	\begin{equation}
		\begin{aligned}
			\dot{V} & \leq -2\mathfrak{e}_{\min}(\varGamma\mathcal{M})V + \frac{\tilde{\varTheta}^{\top}\varGamma^2\tilde{\varTheta}}{2\eta} + \frac{\eta\varLambda^{\top}\varLambda}{2} \\
			        & = -\nu V + \Xi
		\end{aligned}
	\end{equation}
	where $\nu = 2\mathfrak{e}_{\min}(\varGamma\mathcal{M}) - \mathfrak{e}_{\max}(\varGamma^2)/(2\eta)$ and $\Xi = \eta\varLambda^{\top}\varLambda/2$.
	According to the discussions given in the proof of Theorem~\eqref{main}, we know that $2\mathfrak{e}_{\min}(\varGamma\mathcal{M}) > 0$ , so there always exists $\eta>0$ to retain $\nu>0$. 
	Therefore, applying the comparison lemma \cite{khalilNonlinearSystems2002}, we can conclude that the estimation error $\tilde{\varTheta}$ will converge to a compact set $\Omega$, which defined as
	\begin{equation}
		\Omega = \Biggl\{ \tilde{\varTheta}: \|\tilde{\varTheta}\|^2 \leq 2\|\varGamma\|\Bigl( e^{-\nu t}V(\tilde{\varTheta}_0) + \frac{\Xi}{\nu} \Bigr)\Biggr\}
	\end{equation}
	with $\tilde{\varTheta}_0 =\tilde{\varTheta}(t_0)$.
\end{proof}

\section{Numerical Example}\label{Numerical Example}
Consider a simple mass-spring damper system as \cite{slotineAppliedNonlinearControl1991}:
\begin{equation}
	\ddot{y} + \theta_1\dot{y} + \theta_2 y + \theta_3 y^3 = u
\end{equation}
where $y$ is the displacement, $u$ is the input force, and the estimated unknown parameter is $\varTheta=[\theta_1,$ $\theta_2,\theta_3]^{\top}=[10,20,1]^{\top}$. \footnotemark\footnotetext{In this section, apart from those estimators given in Table~\ref{lpsys}, the estimator reported in \cite{choCompositeModelReference2018} is considered for the purpose of comparison.}

Let $x_1 = y$ and $x_2 = \dot{y}$, then we can rewrite it as
\begin{equation}
	\label{mass-spring}
	\dot{x_2} =  u +  \varPhi^{\top}\varTheta, \; \varPhi^{\top} = [-x_2,-x_1,-x_1^2].
\end{equation}
Furthermore, the selection of estimator parameters is summarized in Table \ref{estpara}.\footnotemark\footnotetext{In this table, $\varGamma_0$ denotes the initial value of the time-varying gain for both the LSE and the RAE, and diag[$\cdot$] denotes a diagonal matrix.}

\begin{table}[htb]
	\centering
	\renewcommand\arraystretch{1.2}
	\caption{Estimator parameters }
	\label{estpara}
	\begin{tabular}{cccc}
		\hline
		Parameter                          & Value                                 & Parameter                  & Value   \\
		\hline
		$\kappa$                           & 0.01                                  & $\hat{\varTheta}$          & 0       \\
		$\varGamma_{\textrm{GBE}}$         & $\operatorname{diag}[10^2,10^2,10^3]$ & $\varGamma_{\textrm{RAE}}$ & $10^4I$ \\
		$\varGamma_{\textrm{\cite{choCompositeModelReference2018}}}$ & $10^4I$                               & $\varGamma_0$              & $10^2I$ \\
		$l_0$                              & 1                                     & $\ell_0$                   & 1       \\
		$\bar{\gamma}$                    & $10^3$                                & $\bar{m}$                 & $10^3$  \\
		\hline
	\end{tabular}
\end{table}

\begin{figure}[htbp]
	\centering
	\begin{tikzpicture}[scale=1]
		% \def\xmin{-4}
		% \def\xmax{4}
		% \def\ymin{-2}
		% \def\ymax{2}
		% \def\step{1}
		% \draw[step=\step,gray,very thin] (\xmin,\ymin) grid (\xmax,\ymax);
		% \foreach \x in {\xmin,...,\xmax}
		% \node[anchor=north] at (\x,\ymin-\step/2) {\x};
		% \foreach \y in {\ymin,...,\ymax}
		% \node[anchor=east] at (\xmin-\step/2,\y) {\y};

		\node[rotate=90] at (-3.75*2,0) {\scriptsize $\log_{10}\|\tilde{\varTheta}\|$};
		\node[rotate=0] at (0,-1.85*2) {\scriptsize Time/$s$};
		\node at (0, 0) {\includegraphics[width=0.8\linewidth]{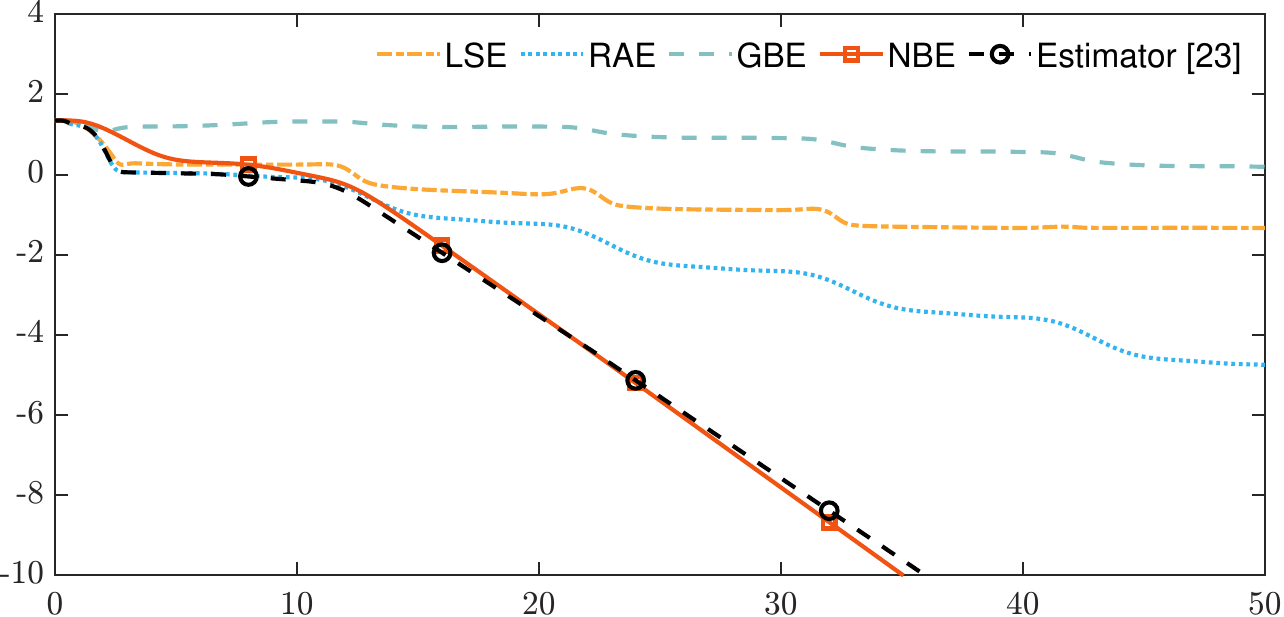}};
	\end{tikzpicture}
	\caption{Comparative results of $\log_{10}{\|\tilde{\varTheta}}\|$ under the PE condition}
	\label{fignorm1}
\end{figure}

\begin{figure}[htbp]
	\centering
	\begin{tikzpicture}[scale=1]
		% \def\xmin{-8}
		% \def\xmax{8}
		% \def\ymin{-8}
		% \def\ymax{8}
		% \def\step{1}
		% \draw[step=\step,gray,very thin] (\xmin,\ymin) grid (\xmax,\ymax);
		% \foreach \x in {\xmin,...,\xmax}
		% \node[anchor=north] at (\x,\ymin-\step/2) {\x};
		% \foreach \y in {\ymin,...,\ymax}
		% \node[anchor=east] at (\xmin-\step/2,\y) {\y};

		\node[rotate=90] at (-4.2*2,0) {\scriptsize $\tilde{\varTheta}$};
		\node[rotate=0] at (0,-3.3*2) {\scriptsize Time/$s$};
		\node at (-4, 4.3) {\includegraphics[width=0.45\linewidth]{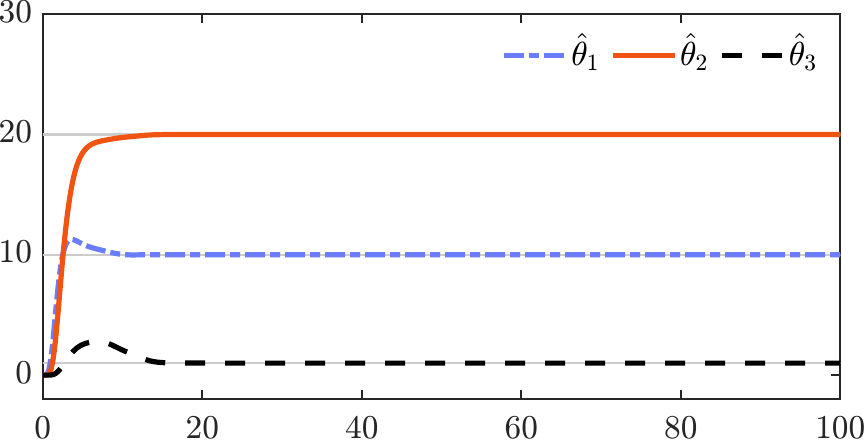}};
		\node at (4, 4.3) {\includegraphics[width=0.45\linewidth]{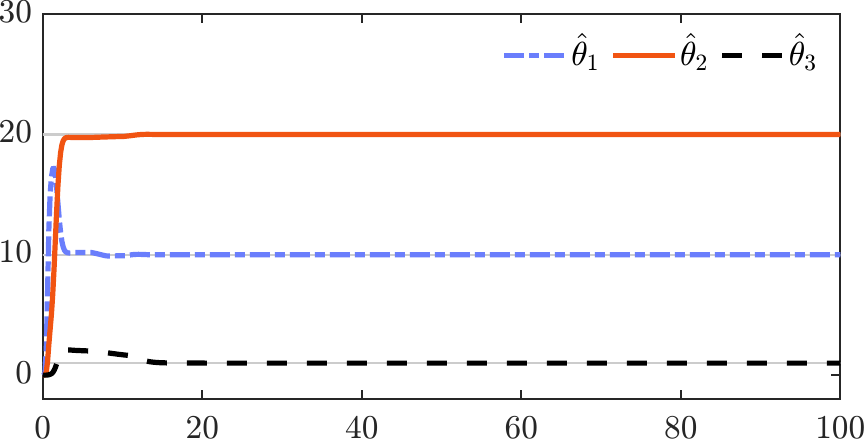}};
		\node at (-4, 0) {\includegraphics[width=0.45\linewidth]{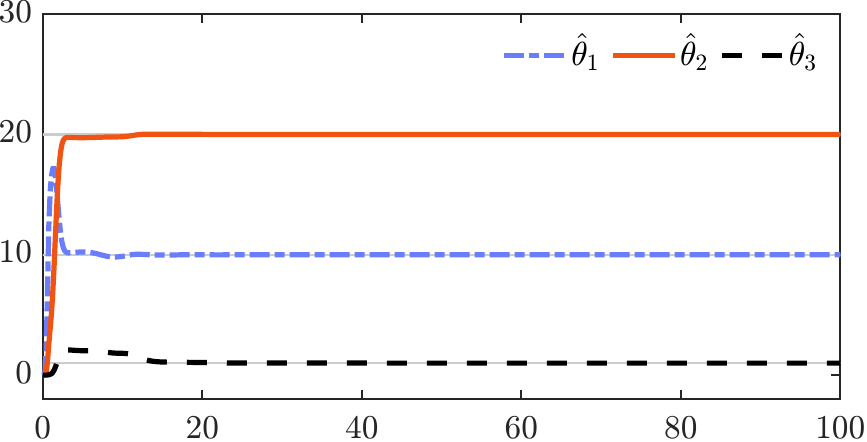}};
		\node at (4, 0) {\includegraphics[width=0.45\linewidth]{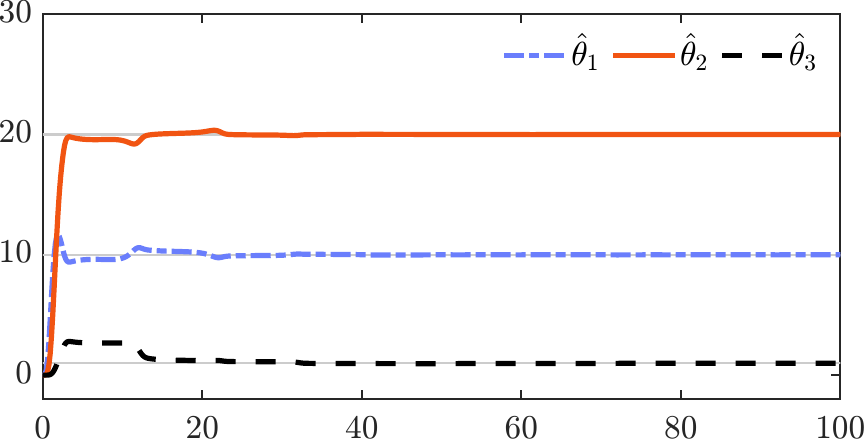}};
		\node at (0, -4.3) {\includegraphics[width=0.45\linewidth]{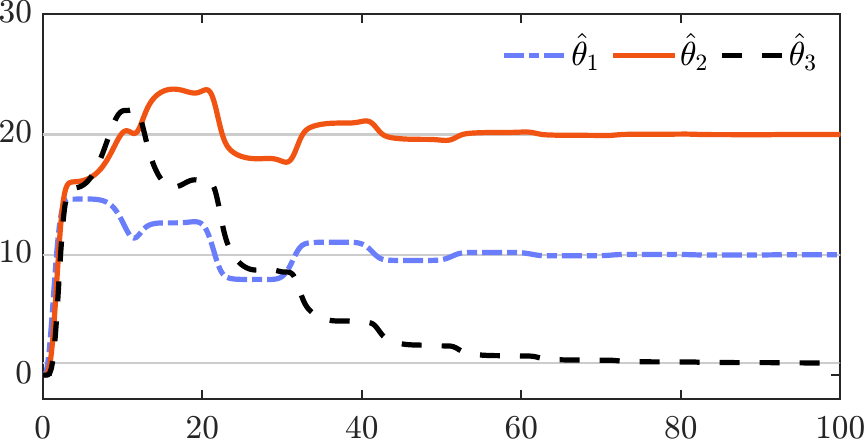}};
		\node[rotate=0] at (-3*2,2.8*2) {\scriptsize NBE};
		\node[rotate=0] at (1.2*2,2.8*2) {\scriptsize Estimator \cite{choCompositeModelReference2018}};
		\node[rotate=0] at (-3*2,0.7*2) {\scriptsize RAE};
		\node[rotate=0] at (1*2,0.7*2) {\scriptsize LSE};
		\node[rotate=0] at (-1*2,-1.35*2) {\scriptsize GBE};
	\end{tikzpicture}
	\caption{Comparative results of $\tilde{\varTheta}$ under the PE condition}
	\label{figpara1}
\end{figure}

\begin{figure}[htbp]
	\centering
	\begin{tikzpicture}[scale=1]
		% \def\xmin{-4}
		% \def\xmax{4}
		% \def\ymin{-2}
		% \def\ymax{2}
		% \def\step{1}
		% \draw[step=\step,gray,very thin] (\xmin,\ymin) grid (\xmax,\ymax);
		% \foreach \x in {\xmin,...,\xmax}
		% \node[anchor=north] at (\x,\ymin-\step/2) {\x};
		% \foreach \y in {\ymin,...,\ymax}
		% \node[anchor=east] at (\xmin-\step/2,\y) {\y};

		\node[rotate=90] at (-3.75*2,0) {\scriptsize $\log_{10}\|\tilde{\varTheta}\|$};
		\node[rotate=0] at (0,-1.85*2) {\scriptsize Time/$s$};
		\node at (0, 0) {\includegraphics[width=0.8\linewidth]{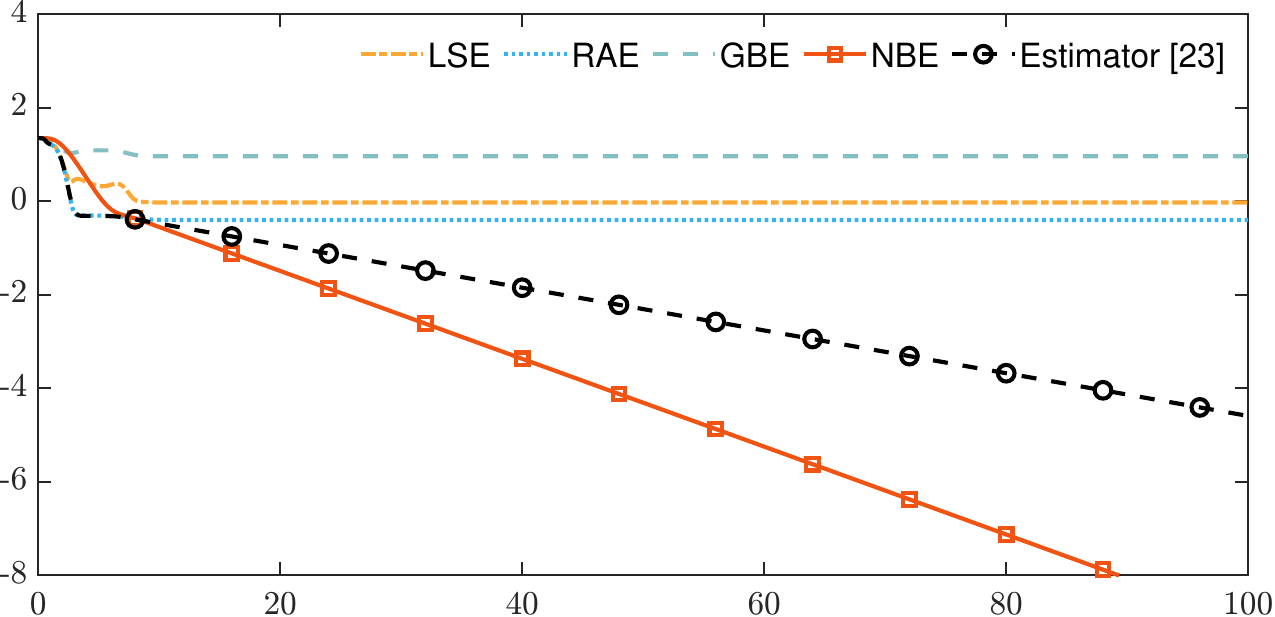}};
	\end{tikzpicture}
	\caption{Comparative results of $\log_{10}{\|\tilde{\varTheta}}\|$ under the FE condition}
	\label{fignorm2}
\end{figure}

\begin{figure}[htbp]
	\centering
	\begin{tikzpicture}[scale=1]
		% \def\xmin{-4}
		% \def\xmax{4}
		% \def\ymin{-4}
		% \def\ymax{4}
		% \def\step{1}
		% \draw[step=\step,gray,very thin] (\xmin,\ymin) grid (\xmax,\ymax);
		% \foreach \x in {\xmin,...,\xmax}
		% \node[anchor=north] at (\x,\ymin-\step/2) {\x};
		% \foreach \y in {\ymin,...,\ymax}
		% \node[anchor=east] at (\xmin-\step/2,\y) {\y};

		\node[rotate=90] at (-4.2*2,0) {\scriptsize $\tilde{\varTheta}$};
		\node[rotate=0] at (0,-3.3*2) {\scriptsize Time/$s$};
		\node at (-4, 4.3) {\includegraphics[width=0.45\linewidth]{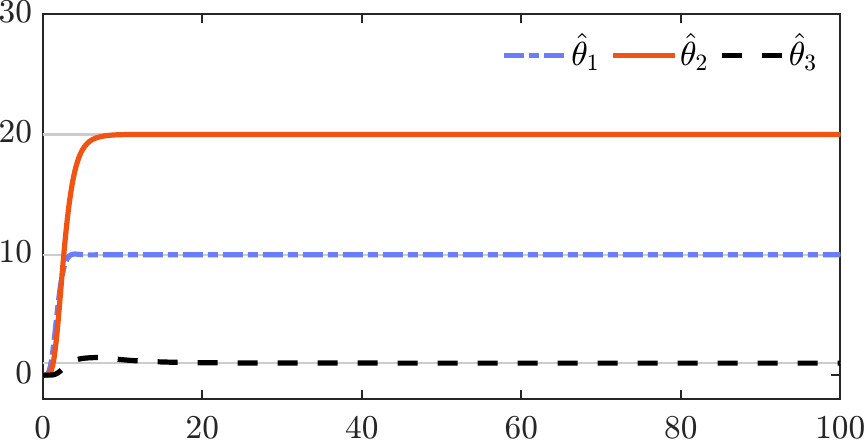}};
		\node at (4, 4.3) {\includegraphics[width=0.45\linewidth]{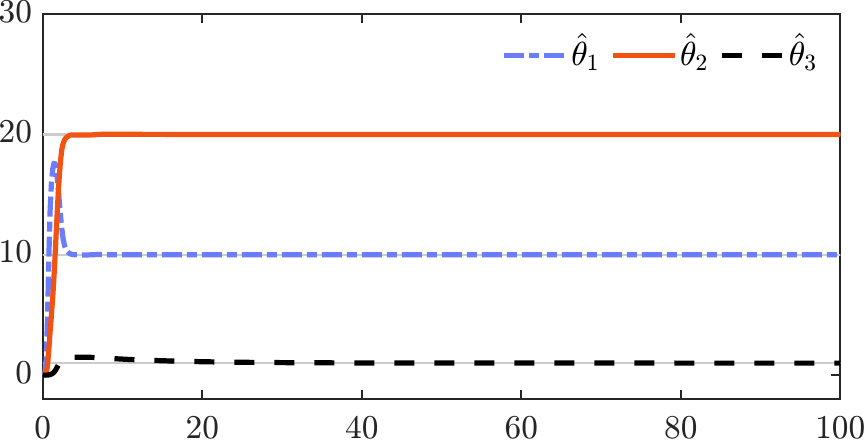}};
		\node at (-4, 0) {\includegraphics[width=0.45\linewidth]{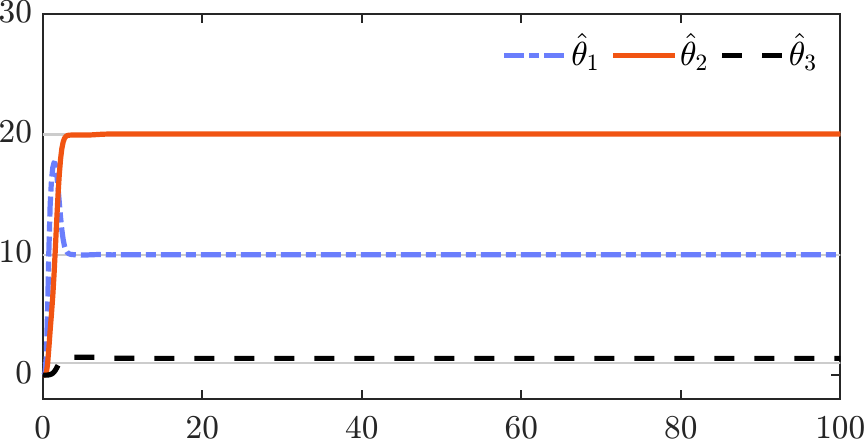}};
		\node at (4, 0) {\includegraphics[width=0.45\linewidth]{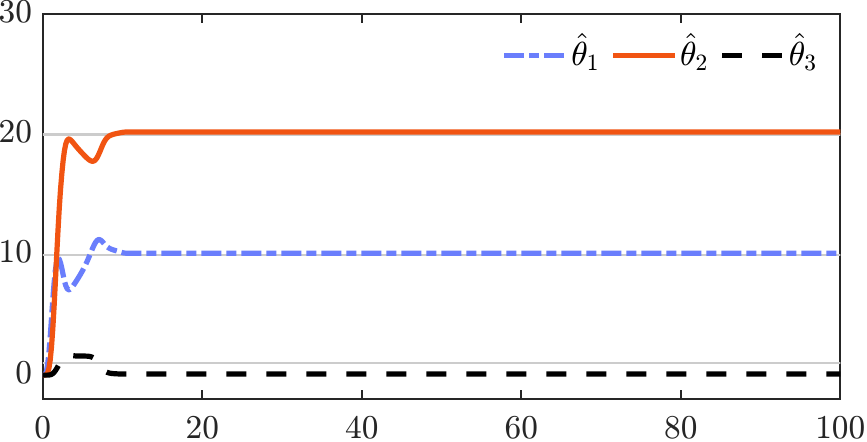}};
		\node at (0, -4.3) {\includegraphics[width=0.45\linewidth]{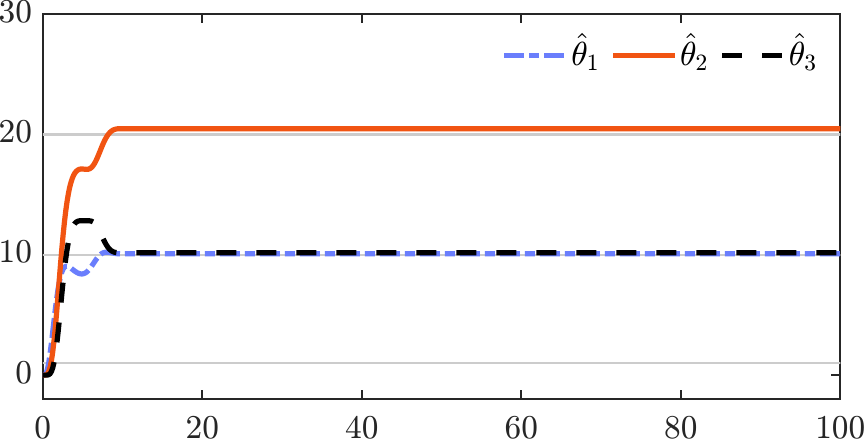}};
		\node[rotate=0] at (-3*2,2.8*2) {\scriptsize NBE};
		\node[rotate=0] at (1.2*2,2.8*2) {\scriptsize Estimator \cite{choCompositeModelReference2018}};
		\node[rotate=0] at (-3*2,0.7*2) {\scriptsize RAE};
		\node[rotate=0] at (1*2,0.7*2) {\scriptsize LSE};
		\node[rotate=0] at (-1*2,-1.35*2) {\scriptsize GBE};
	\end{tikzpicture}
	\caption{Comparative results of $\tilde{\varTheta}$ under the FE condition}
	\label{figpara2}
\end{figure}

\begin{figure}[htbp]
	\centering
	\begin{tikzpicture}[scale=1]
		% \def\xmin{-4}
		% \def\xmax{4}
		% \def\ymin{-2}
		% \def\ymax{2}
		% \def\step{1}
		% \draw[step=\step,gray,very thin] (\xmin,\ymin) grid (\xmax,\ymax);
		% \foreach \x in {\xmin,...,\xmax}
		% \node[anchor=north] at (\x,\ymin-\step/2) {\x};
		% \foreach \y in {\ymin,...,\ymax}
		% \node[anchor=east] at (\xmin-\step/2,\y) {\y};

		\node[rotate=90] at (-3.75*2,0) {\scriptsize $\log_{10}\|\tilde{\varTheta}\|$};
		\node[rotate=0] at (0,-1.85*2) {\scriptsize Time/$s$};
		\node at (0, 0) {\includegraphics[width=0.8\linewidth]{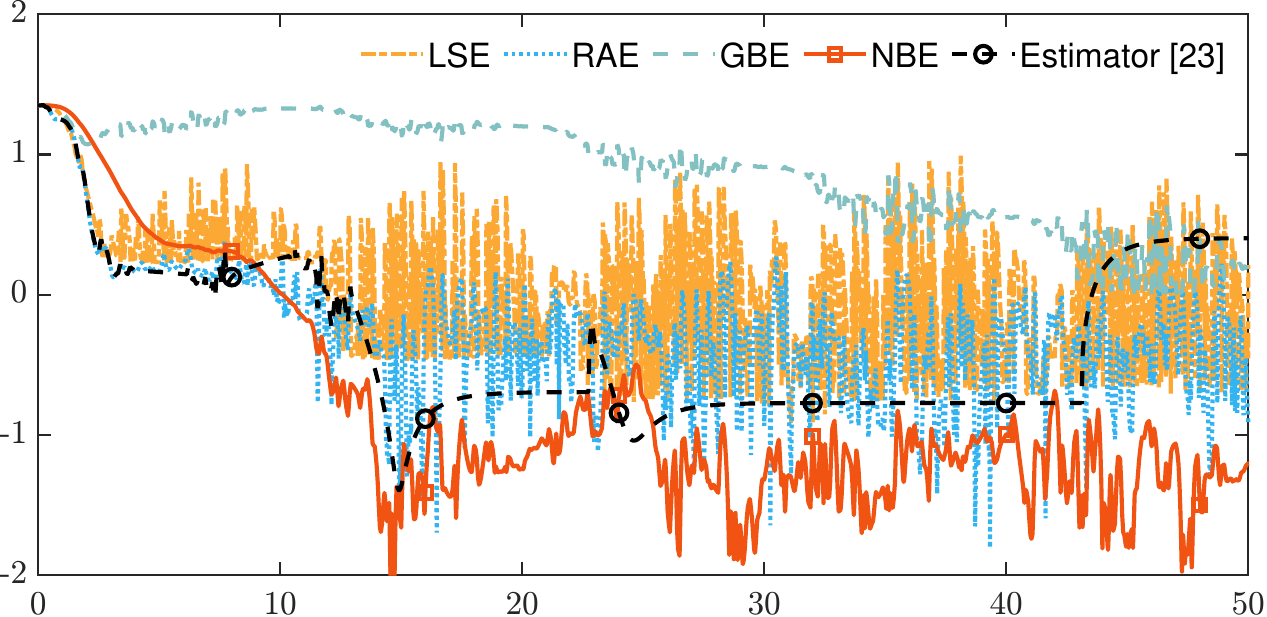}};
	\end{tikzpicture}
	\caption{Comparative results of $\log_{10}{\|\tilde{\varTheta}}\|$ under the PE condition in presence of disturbance}
	\label{fignorm3}
\end{figure}

\begin{figure}[htbp]
	\centering
	\begin{tikzpicture}[scale=1]
		% \def\xmin{-4}
		% \def\xmax{4}
		% \def\ymin{-4}
		% \def\ymax{4}
		% \def\step{1}
		% \draw[step=\step,gray,very thin] (\xmin,\ymin) grid (\xmax,\ymax);
		% \foreach \x in {\xmin,...,\xmax}
		% \node[anchor=north] at (\x,\ymin-\step/2) {\x};
		% \foreach \y in {\ymin,...,\ymax}
		% \node[anchor=east] at (\xmin-\step/2,\y) {\y};

		\node[rotate=90] at (-4.2*2,0) {\scriptsize $\tilde{\varTheta}$};
		\node[rotate=0] at (0,-3.3*2) {\scriptsize Time/$s$};
		\node at (-4, 4.3) {\includegraphics[width=0.45\linewidth]{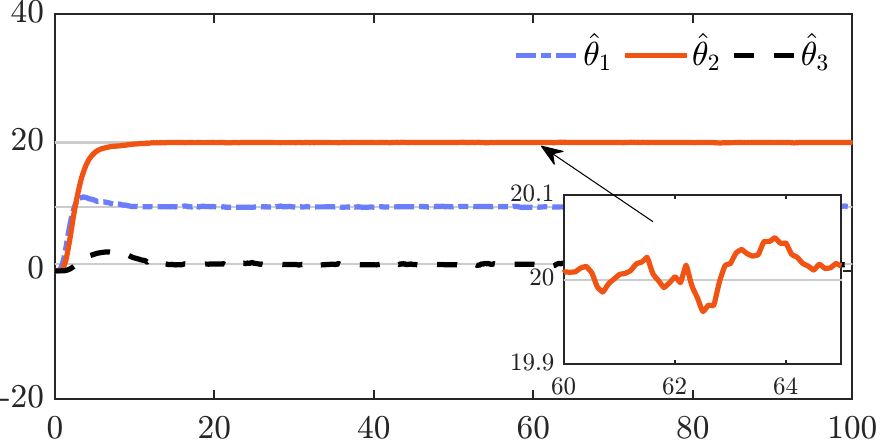}};
		\node at (4, 4.3) {\includegraphics[width=0.45\linewidth]{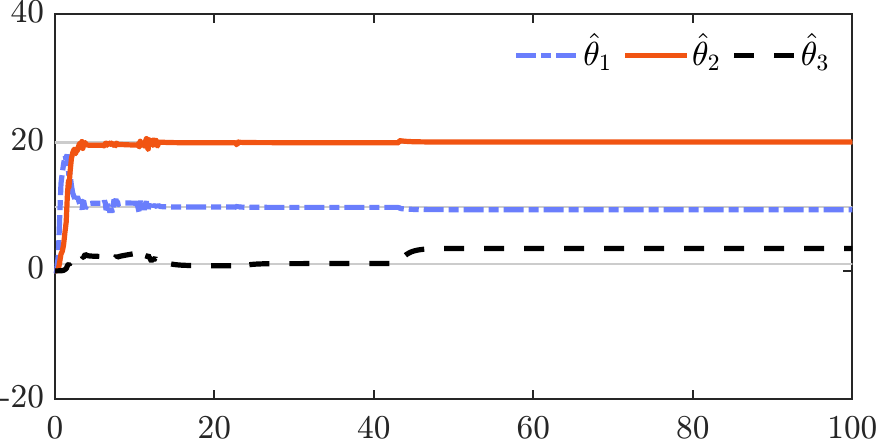}};
		\node at (-4, 0) {\includegraphics[width=0.45\linewidth]{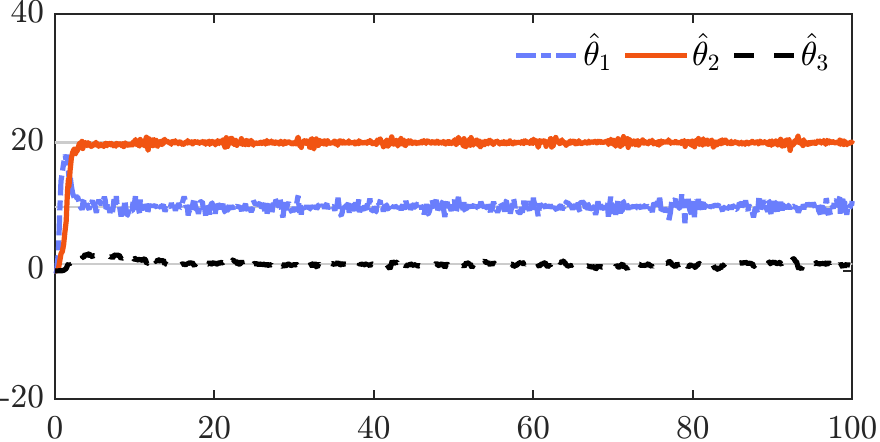}};
		\node at (4, 0) {\includegraphics[width=0.45\linewidth]{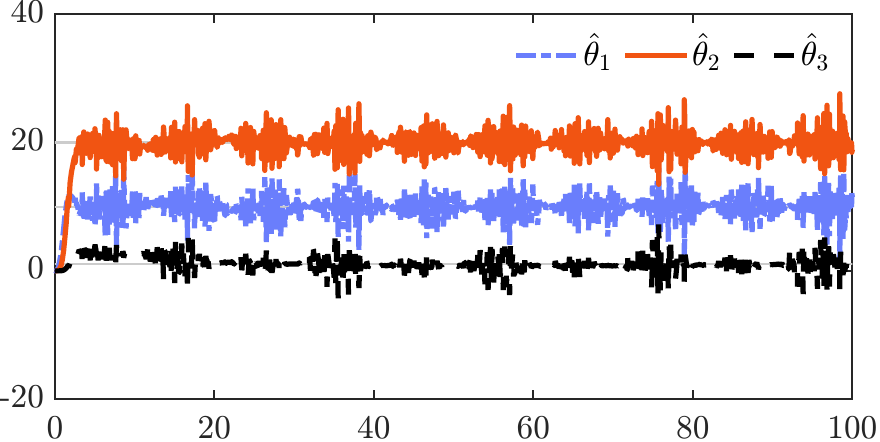}};
		\node at (0, -4.3) {\includegraphics[width=0.45\linewidth]{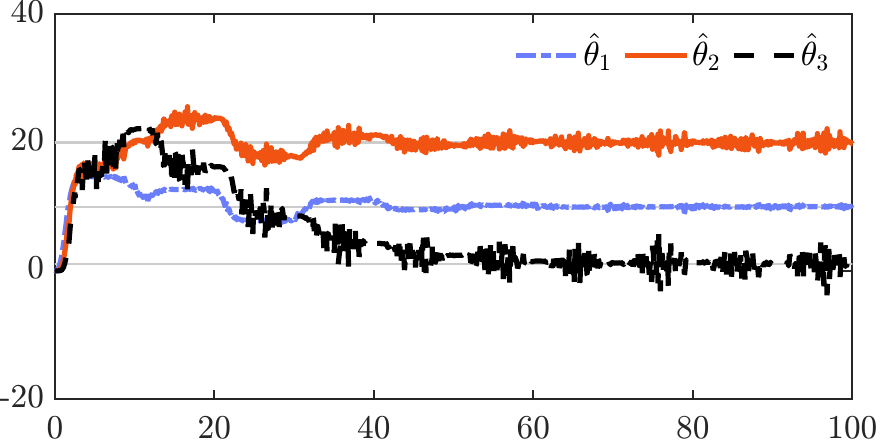}};
		\node[rotate=0] at (-3*2,2.8*2) {\scriptsize NBE};
		\node[rotate=0] at (1.2*2,2.8*2) {\scriptsize Estimator \cite{choCompositeModelReference2018}};
		\node[rotate=0] at (-3*2,0.7*2) {\scriptsize RAE};
		\node[rotate=0] at (1*2,0.7*2) {\scriptsize LSE};
		\node[rotate=0] at (-1*2,-1.35*2) {\scriptsize GBE};
	\end{tikzpicture}
	\caption{Comparative results of $\tilde{\varTheta}$ under the PE condition in presence of disturbance}
	\label{figpara3}
\end{figure}

\begin{figure}[htbp]
	\centering
	\begin{tikzpicture}[scale=1]
		% \def\xmin{-4}
		% \def\xmax{4}
		% \def\ymin{-2}
		% \def\ymax{2}
		% \def\step{1}
		% \draw[step=\step,gray,very thin] (\xmin,\ymin) grid (\xmax,\ymax);
		% \foreach \x in {\xmin,...,\xmax}
		% \node[anchor=north] at (\x,\ymin-\step/2) {\x};
		% \foreach \y in {\ymin,...,\ymax}
		% \node[anchor=east] at (\xmin-\step/2,\y) {\y};

		\node[rotate=90] at (-4.2*2,0) {\scriptsize $\|\varGamma\|$};
		\node[rotate=0] at (0,-1.25) {\scriptsize Time/$s$};
		\node at (-4, 0) {\includegraphics[width=0.45\linewidth]{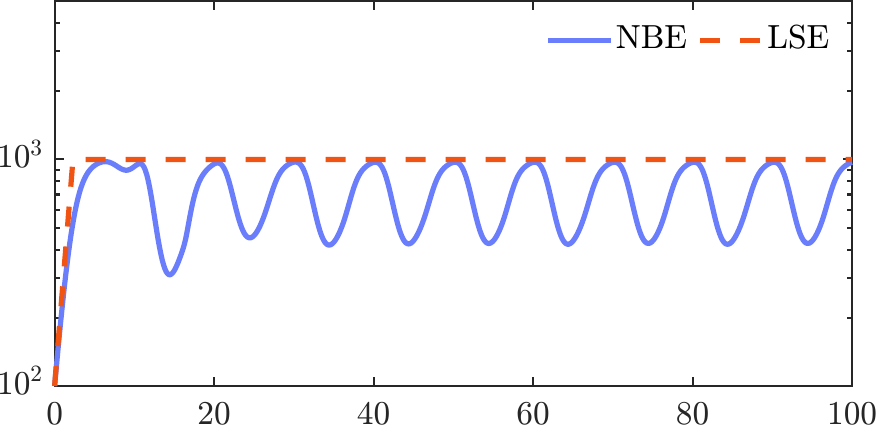}};
		\node at (4, 0) {\includegraphics[width=0.45\linewidth]{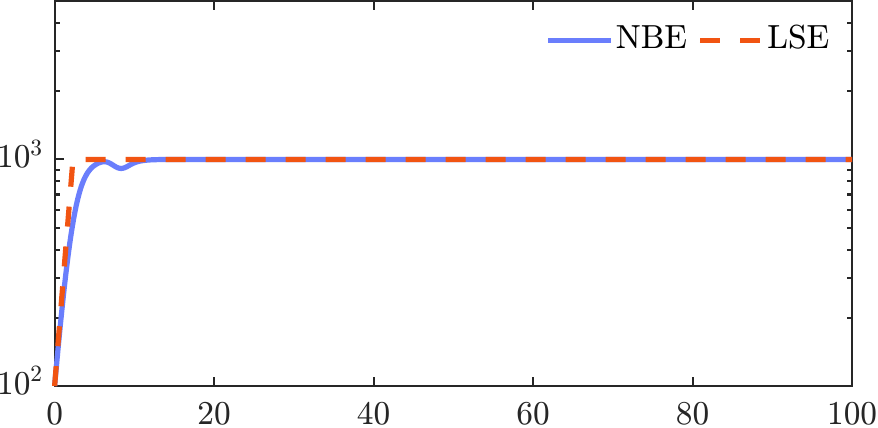}};
	\end{tikzpicture}
	\caption{Comparative results of $\|\varGamma\|$ in presence of disturbance (The left is under the PE condition, and the right is under the FE condition.)}
	\label{figgain}
\end{figure}

\begin{figure}[htbp]
	\centering
	\begin{tikzpicture}[scale=1]
		% \def\xmin{-4}
		% \def\xmax{4}
		% \def\ymin{-2}
		% \def\ymax{2}
		% \def\step{1}
		% \draw[step=\step,gray,very thin] (\xmin,\ymin) grid (\xmax,\ymax);
		% \foreach \x in {\xmin,...,\xmax}
		% \node[anchor=north] at (\x,\ymin-\step/2) {\x};
		% \foreach \y in {\ymin,...,\ymax}
		% \node[anchor=east] at (\xmin-\step/2,\y) {\y};

		\node[rotate=90] at (-3.75*2,0) {\scriptsize $\log_{10}\|\tilde{\varTheta}\|$};
		\node[rotate=0] at (0,-1.85*2) {\scriptsize Time/$s$};
		\node at (0, 0) {\includegraphics[width=0.8\linewidth]{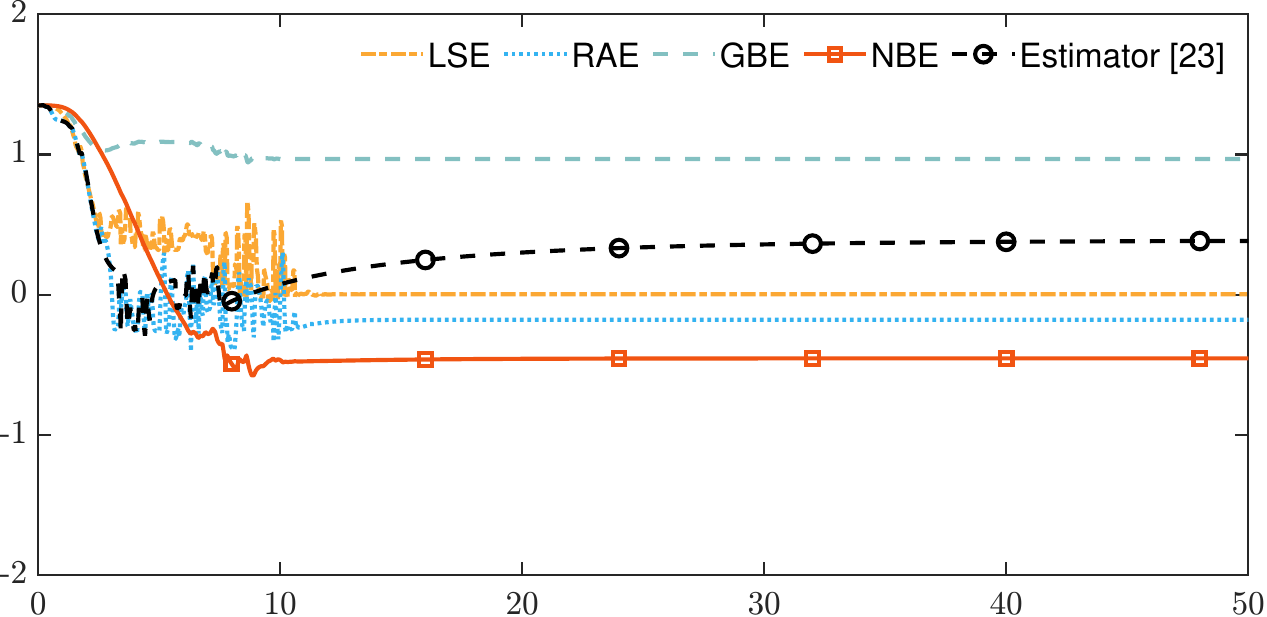}};
	\end{tikzpicture}
	\caption{Comparative results of $\log_{10}{\|\tilde{\varTheta}}\|$ under the FE condition in presence of disturbance}
	\label{fignorm4}
\end{figure}

\begin{figure}[htbp]
	\centering
	\begin{tikzpicture}[scale=1]
		% \def\xmin{-4}
		% \def\xmax{4}
		% \def\ymin{-4}
		% \def\ymax{4}
		% \def\step{1}
		% \draw[step=\step,gray,very thin] (\xmin,\ymin) grid (\xmax,\ymax);
		% \foreach \x in {\xmin,...,\xmax}
		% \node[anchor=north] at (\x,\ymin-\step/2) {\x};
		% \foreach \y in {\ymin,...,\ymax}
		% \node[anchor=east] at (\xmin-\step/2,\y) {\y};

		\node[rotate=90] at (-4.2*2,0) {\scriptsize $\tilde{\varTheta}$};
		\node[rotate=0] at (0,-3.3*2) {\scriptsize Time/$s$};
		\node at (-4, 4.3) {\includegraphics[width=0.45\linewidth]{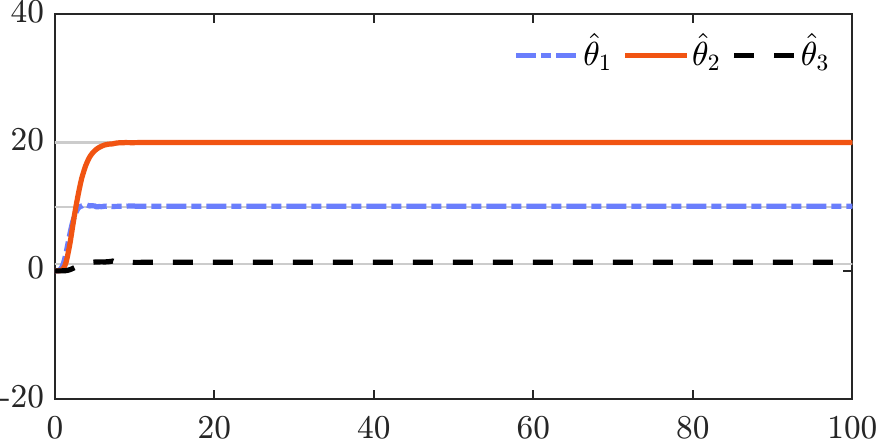}};
		\node at (4, 4.3) {\includegraphics[width=0.45\linewidth]{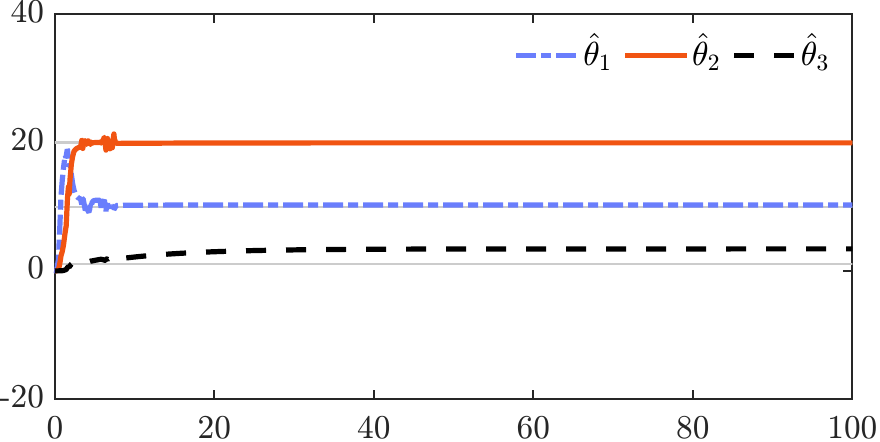}};
		\node at (-4, 0) {\includegraphics[width=0.45\linewidth]{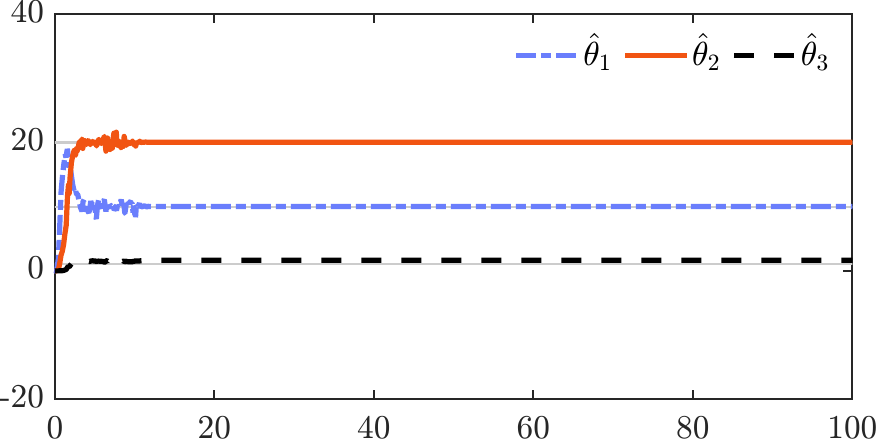}};
		\node at (4, 0) {\includegraphics[width=0.45\linewidth]{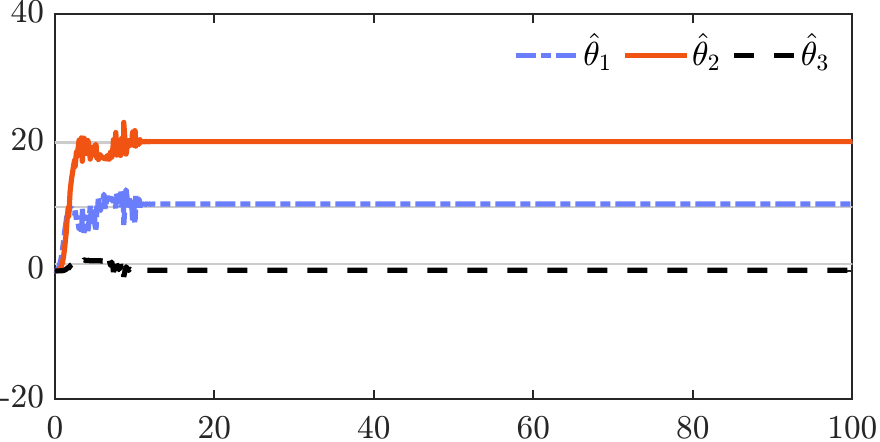}};
		\node at (0, -4.3) {\includegraphics[width=0.45\linewidth]{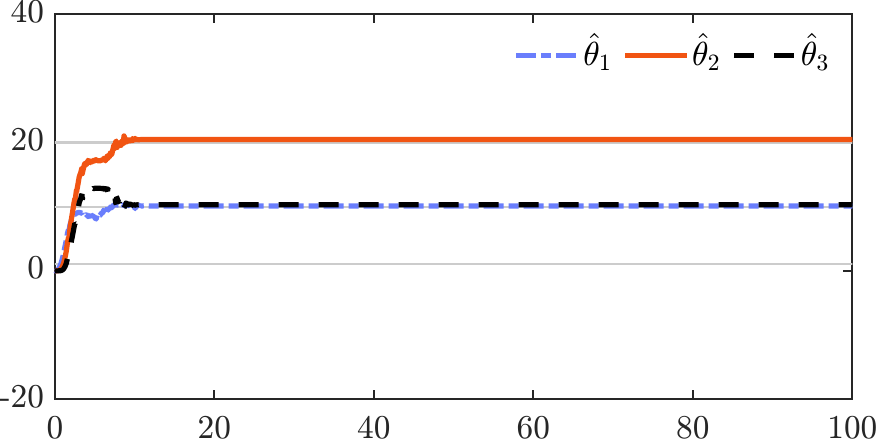}};
		\node[rotate=0] at (-3*2,2.8*2) {\scriptsize NBE};
		\node[rotate=0] at (1.2*2,2.8*2) {\scriptsize Estimator \cite{choCompositeModelReference2018}};
		\node[rotate=0] at (-3*2,0.7*2) {\scriptsize RAE};
		\node[rotate=0] at (1*2,0.7*2) {\scriptsize LSE};
		\node[rotate=0] at (-1*2,-1.35*2) {\scriptsize GBE};
	\end{tikzpicture}
	\caption{Comparative results of $\tilde{\varTheta}$ under the FE condition in presence of disturbance}
	\label{figpara4}
\end{figure}
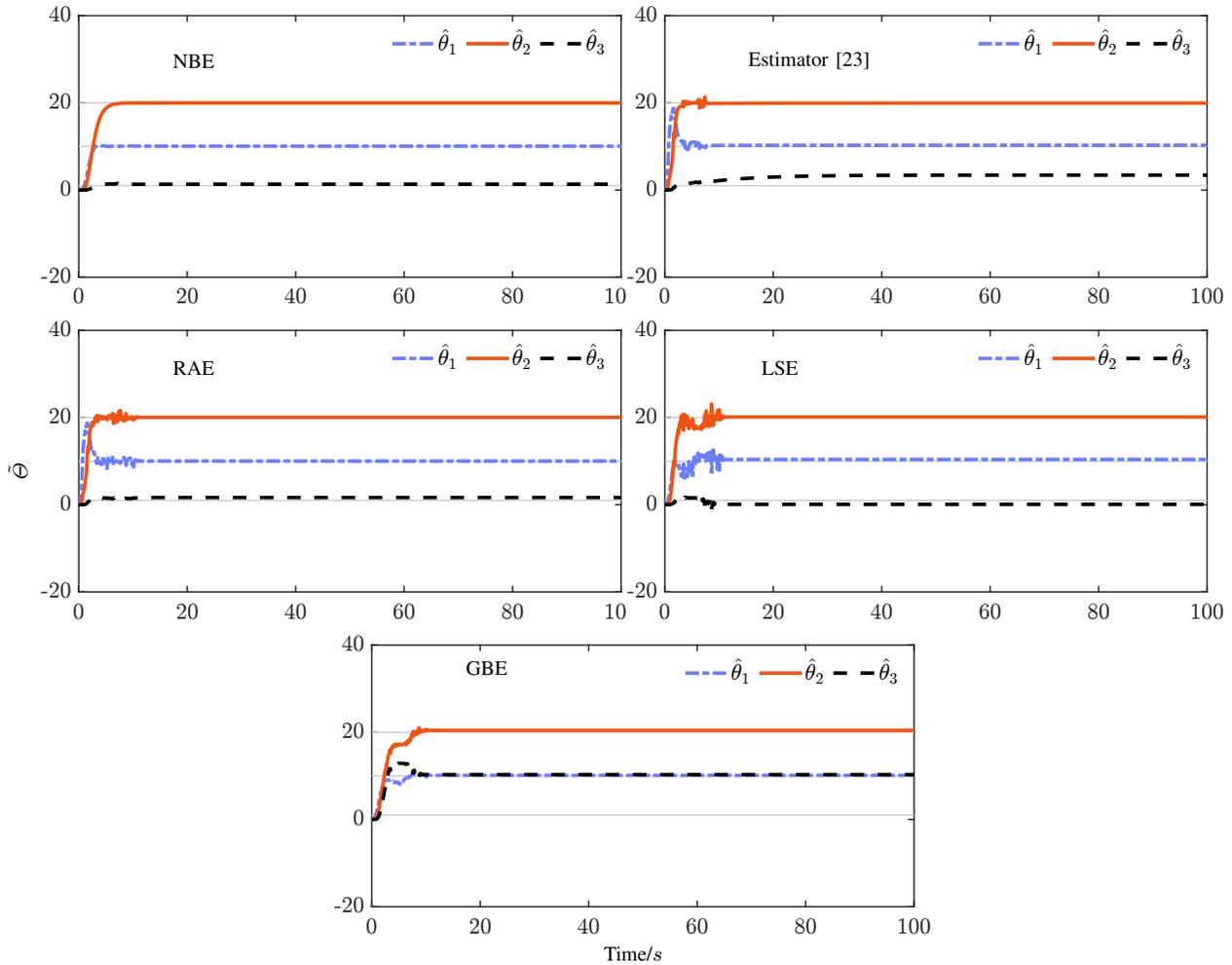

First, the case with the PE condition is demonstrated to verify the effectiveness of all estimators, where the control input $u(t)$ is given as a sinusoidal signal
\begin{equation*}
	u_1 = 5\sin{\frac{\pi}{10}}.
\end{equation*}

The profile of error index $\log_{10}\|\tilde{\varTheta}\| = \log_{10}\|\hat{\varTheta}-\varTheta\|$ is shown in Fig.~\ref{fignorm1}, and the trajectories of the estimated parameter are provided in Fig.~\ref{figpara1}.
Intuitively, accumulating over time, all estimators are able to achieve an accurate convergence, i.e., $\lim\limits_{t\rightarrow\infty}\|\tilde{\varTheta}(t)\|=0$.
With introducing the time-varying gain, it is clear that the NBE is fast in the rate of convergence.
Besides, the Estimator \cite{choCompositeModelReference2018} also has a desired rate as a result of that it has a low-pass filter form, i.e., the application of Kreisselmeier filter such that a higher adaptation gain could be used. 

Next, the case with the FE condition is demonstrated, where the control input $u(t)$ is given as a piecewise continuous signal
\begin{equation*}
	u_2 = \left\{\begin{aligned}
		 & 2\sin{\frac{\pi}{5}}, &  & 0 \leq t \leq 10  \\
		 & 0,                    &  & \textnormal{else}
	\end{aligned}\right. .
\end{equation*}

As given in Fig.~\ref{fignorm2} and Fig.~\ref{figpara2}, in the absence of the PE condition, the GBE, LSE, and RAE cannot ensure the exponential convergence of the parameter estimation. In contrast, the proposed NBE and the Estimator \cite{choCompositeModelReference2018} can retain the ability to achieve accurate parameter estimation with fast, smooth and ultimate convergence, even under this weak excitation.

Furthermore, to test the robustness of these estimators, a bounded disturbance``Band-Limited White Noise'' with power 0.0001 is injected into the system, and the corresponding simulation results can be found in Fig.~\ref{fignorm3}-\ref{figpara4}.
It is observed that the proposed NBE exhibits better convergence properties and robustness.
For the Estimator \cite{choCompositeModelReference2018}, due to the influence of interference on its assessment of the excitation horizon, instances may arise where the convergence outcomes diverge from the true value.
In addition, an examination for $\varGamma$ depicted in Fig.~\ref{figgain} reveals that the gains of the other three estimators (i.e., GBE, RAE, Estimator \cite{choCompositeModelReference2018}) are markedly higher compared to the NBE and the LSE, which would cause an exacerbation of high-frequency oscillation, seeing Fig.~\ref{figpara3} and~\ref{figpara4}.

\section{Conclusion}\label{Conclusion}
In this paper, a novel APE is presented and studied for unknown constant parameters.
Different from the traditional APE schemes that exponentially converge under the PE condition but boundedly converge under the FE condition, the proposed one can achieve an exponential convergence even though in the facing of the FE condition. Besides, another contribution is that a time-varying forgetting factor and a time-varying weight factor are introduced to guarantee the boundedness of APE. The rigorous analysis of stability and robustness is provided via the Lyapunov theorem.
Furthermore, compared to other APE schemes including such a method that is also able to achieve exponential convergence under the FE condition, numerical simulations are given to verify the superiority of the proposed methods.
In future works, we intend to get rid of the dependency of APE on full system states, turning to a more general scenario where only the system output is accessible.

\bibliographystyle{ieeetr}
\bibliography{Reference}

\end{document}